\documentclass[fleqn,usenatbib]{mnras}

\usepackage{newtxtext,newtxmath}

\usepackage[T1]{fontenc}

\DeclareRobustCommand{\VAN}[3]{#2}
\let\VANthebibliography\thebibliography
\def\thebibliography{\DeclareRobustCommand{\VAN}[3]{##3}\VANthebibliography}

\usepackage{tabularx}
\usepackage{multirow}
\usepackage{hyperref}
\usepackage{makecell}
\usepackage{xcolor}
\usepackage{physics}
\usepackage[normalem]{ulem}

\usepackage{graphicx}	
\usepackage{amsmath}	

\newcolumntype{Y}{>{\centering\arraybackslash}X}
\newcolumntype{W}[1]{>{\centering\arraybackslash}p{#1}}

\title[Evolution mapping for the halo mass function]{Evolution mapping III: A new recipe for the halo mass function}

\author[A. Fiorilli et al.]{
Andrea Fiorilli,$^{1}$\thanks{E-mail: fiorilla@mpe.mpg.de (AF)}
Andrés N. Ruiz,$^{2,3}$
Ariel G. Sánchez,$^{1,4}$
and Matteo Esposito$^{1}$
\\
$^{1}$Max-Planck-Institut f\"ur extraterrestrische Physik, Postfach 1312, Giessenbachstr., 85748 Garching, Germany\\
$^{2}$Instituto de Astronomía Teórica y Experimental (CONICET-UNC), Laprida 854, X5000BGR, Córdoba, Argentina\\
$^{3}$Observatorio Astronómico, Universidad Nacional de Córdoba, Laprida 854, X5000BGR, Córdoba, Argentina\\
$^{4}$Universit\"ats-Sternwarte M\"unchen,  Fakult\"at f\"ur Physik, Ludwig-Maximilians-Universit\"at M\"unchen, Scheinerstrasse 1, 81679 M\"unchen, Germany\\
}
\date{Accepted XXX. Received YYY; in original form ZZZ}

\pubyear{\the\year{}}

\begin{document}
\label{firstpage}
\pagerange{\pageref{firstpage}--\pageref{lastpage}}
\maketitle

\begin{abstract}
We present a new prescription for the halo mass function (HMF) built upon the Evolution Mapping framework. This approach provides a physical motivation to parametrise the non-universality of the HMF in terms of the recent history of structure formation and the local shape of the linear matter power spectrum. 
Our model was calibrated against measurements from N-body simulations, with halo samples defined by ten overdensity thresholds, $\Delta$, ranging from 150 to 1600 times the mean background matter density. 
For our reference mass definition, $\Delta=200$, the calibrated fitting function achieves per cent-level accuracy across a wide range of masses, redshifts, and structure formation histories, and maintains this performance when tested on cosmologies with different linear power spectrum shapes. 
This high level of accuracy is maintained across other mass definitions, degrading only slightly to the 5 per cent level at the highest values of $\Delta$. 
We also provide fitting formulae to interpolate the parameters as a function of $\Delta$, which allows for accurate modelling of HMFs defined by intermediate overdensities, with accuracy still well within 5 per cent when tested on halo catalogues defined by the virial overdensity threshold. 
Compared to other commonly used recipes, our prescription yields competitive or superior accuracy across all redshifts and cosmologies, successfully capturing the non-universal features of the HMF where other models exhibit systematic deviations. 
This work provides a high-precision modelling tool for cluster abundance analyses, and demonstrates the power of the evolution mapping framework for building accurate models of observables in the non-linear regime.
\end{abstract}




\section{Introduction}
\label{sec:intro}

Structure formation in the Universe is the result of the growth of small density
fluctuations via gravitational instability, leading to the large-scale structure
pattern known as the cosmic web \citep[e.g., ][]{Peebles_LSS_1980,Bond_LSS_1996}. 
Dark matter haloes form in the peaks of the matter density field and constitute the nodes of the web. They are the first objects to collapse and decouple from the expanding background \citep{press_schechter_hmf_1974,bond_excursion_set_1991}. As the Universe evolves, the formation and merger of small haloes drive their growth in mass across cosmic time, in a way that is
sensitive to the parameters of
the underlying cosmological model \citep{White_Rees_1978,Lacey_Cole_1993,Lacey_Cole_1994}.

Given the high overdensities they attain, dark matter haloes are the preferential sites for the formation of galaxies.
The abundance of haloes as a function of their mass -- the halo mass function (HMF) -- is therefore a fundamental statistic for understanding the distribution of galaxies in the observable Universe. 
In fact, recipes like the halo model \citep[e.g., ][]{cooray_sheth_halo_model_2002, asgari_halo_model_2023} or the halo occupation distribution \citep[HOD, e.g.][]{ berlind_weinberg_hod_2002,zheng_hod_2005} rely explicitly on the HMF to link galaxy clustering statistics to the underlying matter density field, making it essential for extracting cosmological constraints from galaxy clustering observables.
Beyond its role as an ingredient in models of galaxy statistics, the HMF is itself a powerful cosmological observable: the abundance of the most massive haloes directly underlies the use of galaxy cluster number counts as a probe of cosmology \citep[e.g., ][]{sartoris_euclid_clusters_2016, ghirardini_erass1_cosmology_2024}.

In the next years, upcoming and ongoing galaxy surveys like \textit{Euclid}
\citep{laureijs_euclid_red_book_2011, euclid_overview_2024}, the \textit{extended
ROentgen Survey with an Imaging Telescope Array}
\citep[eROSITA,][]{predehl_erosita_instrument_2021}, the Dark Energy Survey
\citep[DES,][]{DES_2021}, the Dark Energy Spectroscopic Instrument
\citep[DESI,][]{DESI_2016,DESI_DR1_2025} and the Legacy Survey of Space
and Time \citep[LSST,][]{LSST_2019}, will generate an unprecedented amount of
data across various wavelengths, including optical, infrared, and X-rays. 
This wealth of data requires the development of increasingly precise statistical tools to accurately
predict the abundance and evolution of dark matter haloes, enabling us to
rigorously test and refine our cosmological models.

Several analytic formalisms have been developed to model the HMF, establishing
the theoretical framework for describing halo abundances as a function of mass \citep{press_schechter_hmf_1974,Peacock_Heavens_1990,bond_excursion_set_1991,
Jedamzik_1995,Monaco_1995,Yano_1996,Audit_1997,corasaniti_achitouv_hmf_2011,Juan_2014,musso_sheth_excursion_2014,musso_sheth_energy_peaks_2021}.
While these analytic methods provide a solid theoretical basis and valuable physical insight,
cosmological N-body simulations have proven to be more effective at capturing the non-linear processes of halo formation, thus allowing precise fits to the HMF across various cosmological models and mass definitions
\citep{Efstathiou_1988,Lacey_Cole_1994,jenkins_hmf_2001,
sheth_mo_tormen_hmf_2001, Evrad_2002, White_2002, reed_hmf_2007, tinker_hmf_2008,
crocce_hmf_2010, courtin_hmf_2011, bhattacharya_hmf_2011, watson_hmf_2013,
Despali_2016, bocquet_hmf_2016, diemer_hmf_2020, seppi_hmf_2021,
ondaromallea_hmf_2021, castro_euclid_hmf_2023,Gavas_2023,
fiorino_png_hmf_2024}. 
For this reason, fitting halo abundances directly from N-body simulations has
become the standard approach for modelling the HMF in the era of precision
cosmology.   

In this paper, we take advantage of the Evolution Mapping approach
\citep{sanchez_evo_mapping_2022, esposito_evolution_mapping_2024} to identify how the HMF is impacted by the history of structure formation in the Universe and by the shape of the linear power spectrum of density fluctuations, $P_\mathrm{L}(k)$. 
We employ high-resolution cosmological N-body simulations to calibrate a new prescription for the halo multiplicity function, $f(\nu)$.  
Our model reaches per cent-level accuracy across an extended
range of halo masses and redshifts, and over a cosmological parameter space
spanning very different, even extreme values of many cosmological parameters.
We calibrate the model for many different overdensity thresholds and show how to interpolate the parameters to allow usage on halo catalogues defined on any
overdensity value.

This paper is organised as follows. In Section~\ref{sec:modelling} we present
the theoretical background to model the HMF and the Evolution Mapping approach.
In Section~\ref{sec:method} we describe the simulations we used, how we identified haloes, and how we computed the HMF. In Section~\ref{sec:results} we present the results
of our fitting procedure, and in Section~\ref{sec:discussion} the discussion of
our findings.
We draw our conclusions in Section~\ref{sec:conclusions}.

\section{Modelling the Halo Mass Function}
\label{sec:modelling}

\subsection{HMF background theory}
\label{subsec:hmf_background}

Several analytic and numerical approaches have been employed to derive HMF models. The pioneering study of \cite{press_schechter_hmf_1974} employed the spherical collapse model \citep{gunn_gott_spherical_collapse_1972} to compute the number density of haloes in an Einstein-de Sitter (EdS) Universe.
They showed that, under such assumptions, the fractional number density of haloes is uniquely a function of the halo peak height,
\begin{equation}
    \label{eq:peak_height_def}
    \nu \equiv \delta_{\mathrm{c}} / \sigma(M, \, z) \, .
\end{equation}
Here $\delta_{\mathrm{c}}$ is the linear overdensity threshold for spherical collapse, corresponding in an EdS universe to $\delta_{\mathrm{c}} = 3/5 \, (3 \pi / 2)^{2/3} \simeq 1.686$, with sub-per cent corrections in cosmologies with dark energy \citep{kitayama_suto_1996}.
We adopt the EdS value throughout, as the mild non-universality induced by the cosmology dependence of $\delta_\mathrm{c}$ is effectively absorbed by the other parameters of our model.
The denominator $\sigma(M, \, z)$ is the RMS fluctuation of the linear density field at redshift $z$, with the field smoothed over a top-hat filter of the same scale as the Lagrangian radius of the halo, $R_{\mathrm{L}} = {(3M / 4\pi \bar{\rho})}^{1/3}$, $\bar{\rho}$ being the mean comoving matter density.

The abundance of dark matter haloes is commonly expressed in terms of the multiplicity function, $f(\nu)$, which accounts for the amount of matter that has collapsed
into haloes and is related to the number density of haloes of mass $M$ at redshift $z$ by
\begin{equation}
    \label{eq:fnu_dndm}
     \dv{n}{\ln M} = \frac{\bar{\rho}}{M} f(\nu) \dv{\ln \nu}{\ln M} \, .
\end{equation}
The property of all cosmological and redshift dependence being encoded only through $\sigma(M, \, z)$, while $f(\nu)$ is a single cosmology-independent function, is referred to as the universality of the mass function. Conversely, any explicit dependence of $f(\nu)$ on time or cosmology not captured by $\sigma(M, \, z)$ is termed non-universality.

Assuming that the collapsing region is isolated and spherically symmetric, the  multiplicity function is given by \citep{press_schechter_hmf_1974}
\begin{equation}
    \label{fnu_ps74}
    f_\mathrm{PS74}(\nu) = \sqrt{\frac{2}{\pi}} \, \nu \, \exp\left(-\frac{1}{2} \nu^2\right) \, .
\end{equation}
This formalism was later refined by \cite{sheth_mo_tormen_hmf_2001}, who dropped the assumption of spherical symmetry and used the excursion set approach \citep{bond_excursion_set_1991} to derive an HMF for ellipsoidal collapse, providing a more accurate analytic fit to N-body simulations.

Even though the first theoretical studies derived a universal $f(\nu)$, N-body simulation results have shown that there is a dependency not only on cosmology and redshift, but also on halo definition and the overdensity threshold used.
Since then, high-resolution N-body simulations have become indispensable for calibrating the HMF with higher accuracy. Several fitting functions have been proposed based on the results of simulations, most of them characterising the non-universality of $f(\nu)$ as a scaling of the fitting parameters with redshift.
Examples of these semi-analytic prescriptions include \citet{jenkins_hmf_2001, warren_hmf_2006, reed_hmf_2007, crocce_hmf_2010, courtin_hmf_2011, bhattacharya_hmf_2011, watson_hmf_2013, bocquet_hmf_2016, diemer_hmf_2020, seppi_hmf_2021, ondaromallea_hmf_2021, castro_euclid_hmf_2023, verza_hmf_vsf_2024}, and the widely used formulation by \citet{tinker_hmf_2008}, in which the multiplicity function takes the form
\begin{equation}
    f_\mathrm{T08}(\nu) = A \left(1 + (b \nu)^{a} \right) \exp \left(-c \nu^2\right) \, ,
    \label{eq:T08_fnu}
\end{equation}
where the four parameters $A,\, a,\, b$ and $c$ were fitted to the results of N-body simulations, and have a scaling with redshift, highlighting a significant deviation from universality.
Variants and extensions have also been calibrated for modified theories of gravity \citep[e.g., ][]{gupta_hmf_modified_gravity_2024}, as well as for cosmologies with massive neutrinos \citep{costanzi_hmf_neutrinos_2013, castro_euclid_hmf_2023} or primordial non-Gaussianity \citep{matarrese_hmf_nongaussianity_2000, loverde_hmf_nongaussianity_2008, d'amico_hmf_nongaussianity_2011, fiorino_png_hmf_2024}. 

More recently, emulators, machine learning models trained on extensive suites of N-body simulations, have also emerged as tools for predicting the HMF by interpolating it as a function of cosmological parameters
\citep{mcclintock_aemulus_hmf_emulator_2019, bocquet_miratitan_hmf_emulator_2020, saez_emantis_hmf_emulator_2024, guo_aemulus_hmf_autoencoder_2024, buisman_nn_hmf_2025}. 
While these emulators provide accurate predictions, our goal is to develop a more physically motivated model for the non-universality of the HMF based on the evolution mapping approach. The remainder of this section is devoted to describing this framework.

\subsection{Evolution Mapping and the dependence on the history of structure formation}

The HMF is known to depend on the growth of structure \citep{ondaromallea_hmf_2021, castro_euclid_hmf_2023}. In this work, we aim to refine the modelling of this dependence by exploiting the Evolution Mapping framework \citep{sanchez_evo_mapping_2022, esposito_evolution_mapping_2024}. 
This approach is built on a set of cosmological parameters that excludes any explicit dependence on the dimensionless Hubble parameter, $h$. This restriction allows the parameters to be classified according to the effect they have on the linear matter power spectrum, $P_\mathrm{L}(k)$, into two categories:
\begin{itemize}
    \item \textbf{Shape parameters}, which affect the shape of $P_{\mathrm{L}}(k)$. These include the physical baryon and cold dark matter densities, $\omega_{\mathrm{b}}$ and $\omega_{\mathrm{c}}$, and the primordial scalar spectral index, $n_{\mathrm{s}}$;
    \item \textbf{Evolution parameters}, whose only effect is to scale the amplitude of the linear power spectrum. This group contains, e.g., the physical dark energy density, $\omega_\mathrm{DE}$, 
    its equation of state parameters,
    the curvature energy density $\omega_K$ or the primordial power spectrum amplitude $A_\mathrm{s}$. 
\end{itemize}
Thus, all evolution parameters exhibit a perfect degeneracy at the linear level and can therefore be encapsulated in a single quantity, representing the amplitude of $P_\mathrm{L}(k)$.
However, this role cannot be fulfilled by $\sigma_{8/h}$, the RMS of the linear density fluctuations smoothed over spheres of radius $R=8\,h^{-1}\mathrm{Mpc}$, due to the dependence of the filtering scale on $h$, which results in a mixing of shape and evolution parameters as
\begin{equation}
    h^2 = \sum_i \omega_i \, .
    \label{eq:hubble_sum}
\end{equation}
We chose the notation $\sigma_{8/h}$, rather than just $\sigma_8$, to highlight such dependence explicitly.

\begin{figure}
    \includegraphics[width=0.95\linewidth]{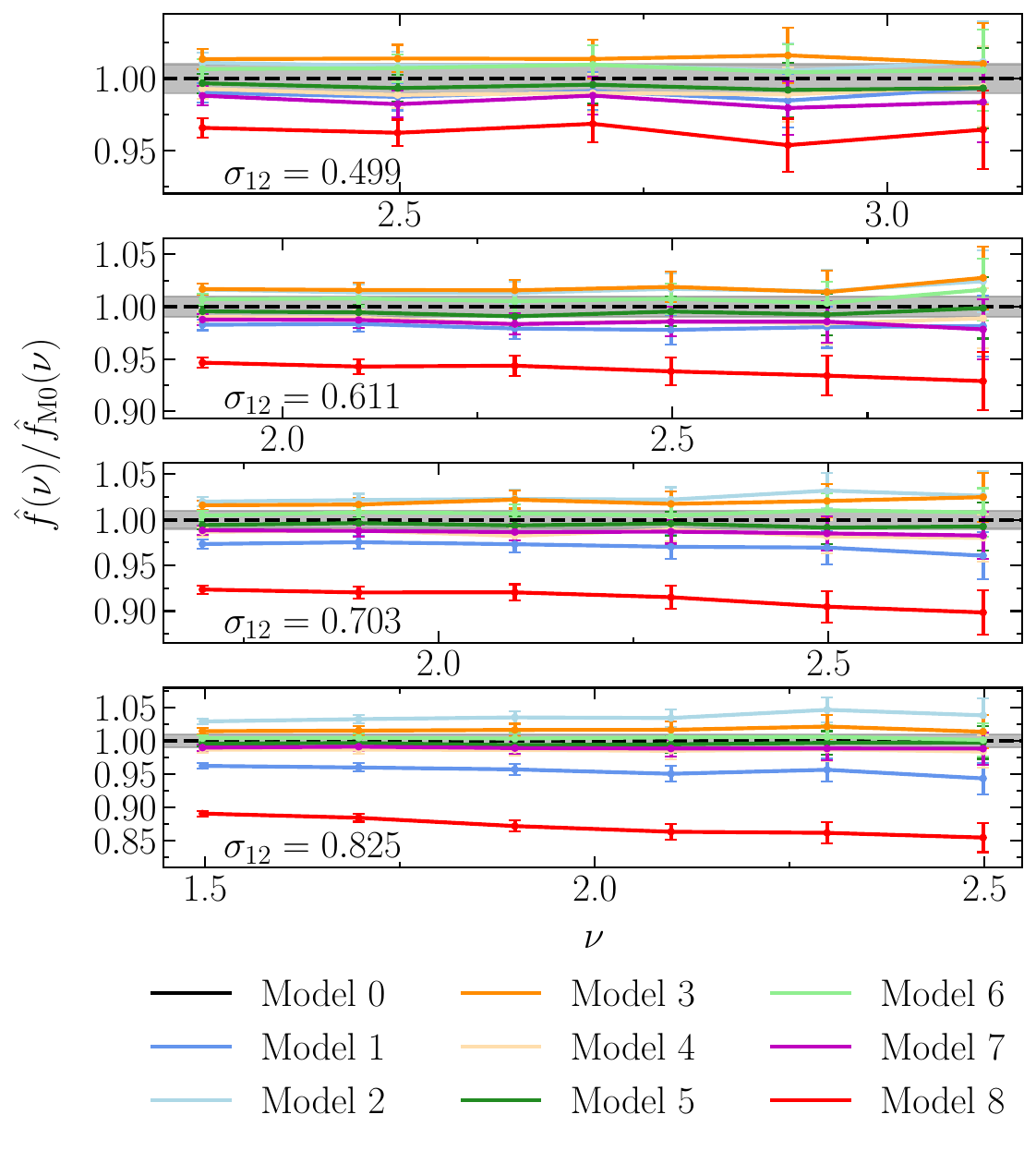}
    \caption{Ratio of the halo multiplicity functions measured in the Aletheia simulations (see Section~\ref{subsec:sims}) to the results from Model 0 at equal clustering amplitude $\sigma_{12}$. The grey bands correspond to a 1 per cent difference.}
    \label{fig:aletheia_ratio_tomodel0}
\end{figure}

Two cosmologies with identical $P_\mathrm{L}(k)$ but different values of $h$ would still be assigned different values of $\sigma_{8/h}$, simply because the smoothing scale itself is $h$-dependent.
This issue was addressed by \citet{sanchez_arguments_against_h_2020}, who advocated for adopting a cosmology-independent comoving scale as a reference to compute the RMS of density fluctuations: choosing a scale of $R = 12 \, \mathrm{Mpc}$\footnote{For a value of $h = 0.67$, $8~ h^{-1}{\rm Mpc} \simeq 12~{\rm Mpc}$.}, the resulting clustering amplitude is defined as $\sigma_{12}$. 

The linear matter power spectrum can then be expressed as 
\begin{equation}
    P_{\mathrm{L}}(k|z,\vb{\Theta}_{\mathrm{s}},\vb{\Theta}_{\mathrm{e}}) = P_{\mathrm{L}}\left(k|\vb{\Theta}_{\mathrm{s}},\sigma_{12}\left(z,\vb{\Theta}_{\mathrm{s}},\vb{\Theta}_{\mathrm{e}}\right)\right).
    \label{eq:pk_evmap_linear}
\end{equation}
Given two cosmologies sharing the set of shape parameters $\vb{\Theta_\mathrm{s}}$, evaluating the linear power spectra at the same $\sigma_{12}$, rather than at the same redshift, gives identical $P_\mathrm{L}(k)$ even for very different $\vb{\Theta_\mathrm{e}}$  \citep[see Fig. 2 of ][]{sanchez_evo_mapping_2022}. \citet{pezzotta_evomapping_neutrinos_2025} showed that the evolution mapping approach can be extended also to cosmologies with massive neutrinos.

\begin{figure}
    \centering
    \includegraphics[width=0.95\linewidth]{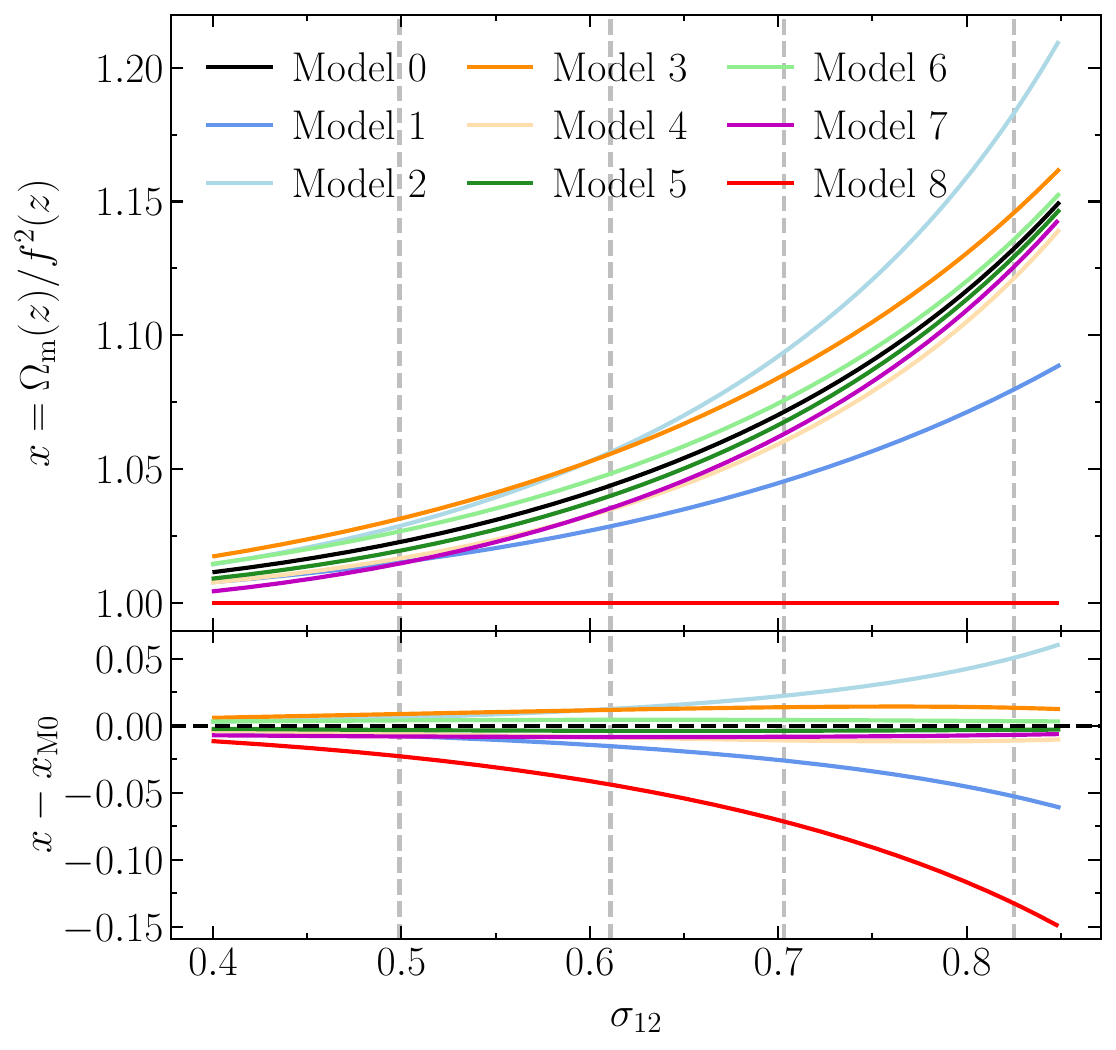}
    \caption{Upper panel: scaling of $x$ as a function of $\sigma_{12}$ in the Aletheia cosmologies defined in Table~\ref{tab:models_aletheia}. Bottom panel: difference between these cosmologies and that of Model 0, $x_\mathrm{M0}$, the reference $\Lambda\mathrm{CDM}$ cosmology. The vertical dashed lines mark the values of $\sigma_{12}$ characterising the simulation snapshots. The differences $x - x_\mathrm{M0}$ at each of the marked values of $\sigma_{12}$ show a remarkable similarity to the multiplicity function ratios shown in the corresponding panel of Fig.~\ref{fig:aletheia_ratio_tomodel0}.}
    \label{fig:x_sigma12_aletheia}
\end{figure}

The non-linear evolution of cosmic structures can be modelled by including higher-order corrections to the perturbative expansion of the matter density field using standard perturbation theory \citep[SPT, see e.g.,][]{bernardeau_pt_2002}, or related approaches such as renormalised perturbation theory \citep{crocce_scoccimarro_renormalized_pt_2006} and the effective field theory of large-scale structure \citep{carrasco_eftoflss_2012, baumann_eftoflss_2012}.
In SPT, the mode-coupling kernels that describe the interactions between Fourier modes of the density field introduce an additional dependence on cosmology, beyond the linear power spectrum $P_\mathrm{L}(k)$. The solutions depend on the ratio
\begin{equation}
x(z) = \frac{\Omega_{\mathrm{m}}(z)}{f^2(z)} \, ,
\label{eq:def_x}
\end{equation}
where $\Omega_{\mathrm{m}}(z)$ is the fractional matter density parameter at redshift $z$ and $f(z) = \mathrm{d}\ln D(z) / \mathrm{d}\ln a$ is the linear growth rate. 

In SPT, such dependence is often neglected, setting $x$ to its value in an Einstein-de Sitter (EdS) universe, $x \equiv 1$. In this case, it is possible to find an analytic solution to the SPT equations without sacrificing much accuracy in the modelling \citep[see, e.g.,][]{Taruya2016, Garny2021}.
Under this approximation, the evolution mapping relation of equation~(\ref{eq:pk_evmap_linear}) would also hold at the non-linear level.

This non-linear behaviour was tested by \citet{sanchez_evo_mapping_2022} using the Aletheia simulations, which were specifically designed to probe the evolution mapping framework. These simulations share identical shape parameters $\vb{\Theta_\mathrm{s}}$, but span a wide range of different evolution parameters $\vb{\Theta_\mathrm{e}}$. By comparing simulation outputs at redshifts chosen to match the same value of $\sigma_{12}$, they showed that the evolution mapping relation of equation~(\ref{eq:pk_evmap_linear}) holds approximately even well into the non-linear regime. The deviations in the power spectra were found to be at the per cent to a few per cent level, even for very different cosmologies. A similar level of agreement was shown to hold for statistics of the velocity field in \cite{esposito_evolution_mapping_2024}.

The residual differences from the exact evolution mapping degeneracy in the non-linear $P(k)$ are driven by the different evolutionary paths these models take to reach the same clustering amplitude, that is, their different integrated growth of structure histories. Recently, \citet{sanchez_aletheia_2025} showed that these non-universal deviations can be accurately modelled by introducing a new parameter, $\tilde{x}$, which integrates the SPT cosmology-dependent variable, $x(z)$, over the recent history as
\begin{equation}
    \tilde{x} (\tau|\eta) = \int_{-\infty}^\tau \mathrm{d}\tau' \, x(\tau') \mathcal{G}(\tau' -\tau | \eta) \, ,
    \label{eq:x_tilde_def}
\end{equation}
where $\tau = \ln \sigma_{12}$ and $\mathcal{G}$ is a one-sided normalised Gaussian kernel. 
The standard deviation $\eta$ controls the memory of the integration.
\citet{sanchez_aletheia_2025} found that using a fixed value of $\eta=0.12$, is sufficient to capture the non-universality in $P(k)$ over a wide range of scales.

This framework has direct implications for the HMF. 
Models following the evolution mapping relation, that is, with identical $\vb{\Theta_\mathrm{s}}$ and $\sigma_{12}$, not only share the same $P_\mathrm{L}(k)$ but also the same linear mass variance, $\sigma(M)$. 
Standard universal HMF models, in which the multiplicity function $f(\nu)$ is cosmology-independent, would therefore predict identical mass functions for all these cosmologies.

However, as in the case of the power spectrum, this universality is not exact. 
While deferring the details of our measurements of the HMF from simulations to Section~\ref{sec:method}, we give an example of this in Fig.~\ref{fig:aletheia_ratio_tomodel0}. 
The HMF measured in snapshots with the same $P_\mathrm{L}(k)$, but different structure formation histories, exhibits clear deviations from perfect universality, with differences that increase with $\sigma_{12}$ and, mildly, with peak height $\nu$.

The strong analogy with the $P(k)$ deviations suggests that $\tilde{x}$ may also describe this non-universality. 
This is supported by Fig.~\ref{fig:x_sigma12_aletheia}, which shows the scaling of $x$ as a function of $\sigma_{12}$ for the Aletheia cosmologies (listed in Table~\ref{tab:models_aletheia}). 
The growing separation between the $x(\tau)$ curves strongly correlates with the growing deviations seen in the HMF ratios in Fig.~\ref{fig:x_sigma12_aletheia}.

Given the ability of $\tilde{x}$ to describe the deviations in $P(k)$ from the perfect evolution mapping relation, we adopt it here to model the non-universality of the multiplicity function.
However, since the HMF probes different, more non-linear scales than the $P(k)$ analysis of \citet{sanchez_aletheia_2025}, we do not assume a fixed memory scale. Instead, we will treat $\eta$ as a free parameter in our model, allowing the HMF data to determine the most relevant memory scale for halo formation.

\subsection{Dependence on the shape of the linear power spectrum}
\label{sec:neff}

The universality of the HMF is also known to be broken by a dependence on the shape of the linear matter power spectrum, $P_\mathrm{L}(k)$ \citep[see, e.g., ][]{jenkins_hmf_2001,warren_hmf_2006, betancortrijo_hmf_redshift_dependence_2006, ondaromallea_hmf_2021, castro_euclid_hmf_2023}. 
The first numerical recipe for $f(\nu)$ with an explicit dependence on the local shape of the power spectrum was provided by \citet{reed_hmf_2007}, and parametrised in terms of the effective spectral index, $n_\mathrm{eff}(\nu)$, which can be defined as a function of peak height as
\begin{equation}
    \label{eq:n_eff_OM}
    n_\mathrm{eff}(\nu) = -2 \dv{\ln \sigma}{\ln R}\Big|_{R_\mathrm{L}(\nu, z)} - 3 \, .
\end{equation}
This quantity measures the slope of the variance $\sigma^2(R)$ at the Lagrangian scale $R_\mathrm{L}(\nu, z)$ corresponding to a peak height $\nu$ at redshift $z$. 
The logarithmic slope of $\sigma(R)$ is better suited for this purpose than the slope of $P_\mathrm{L}(k)$ itself, as the latter exhibits significant oscillations due to the baryonic acoustic oscillations (BAO) signal, but no such feature has ever been detected in the shape of $f(\nu)$. 
The numerical factors appearing in equation~(\ref{eq:n_eff_OM}) are chosen such that for a scale-free power-law matter power spectrum, $P(k) \propto k^n$, one obtains $n_\mathrm{eff}(\nu) = n$.

An explicit dependence on $n_\mathrm{eff}(\nu)$ has become a standard feature in modern HMF prescriptions \citep[e.g., ][]{ondaromallea_hmf_2021,castro_euclid_hmf_2023}. 
In this study, we will also include a similar dependency in our model;
however, we find that the non-universal behaviour of $f(\nu)$ is well captured by the effective slope evaluated at the characteristic peak height of $\nu = 1$, which represents the typical mass of collapsing haloes at the redshift $z$ at which we are evaluating the multiplicity function. 
To simplify our notation, we will hereafter refer to this parameter simply as $n_\mathrm{eff}$, that is
\begin{equation}
\label{eq:n_eff_nu_1}
n_\mathrm{eff} \equiv n_\mathrm{eff}(\nu = 1, z)\,.
\end{equation}
Together with $\tilde{x}$, this parameter completes the set of physical variables we will use to describe the 
non-universality of the HMF.

\subsection{A new recipe for $f(\nu)$}
\label{sec:recipe_fnu}

In the context of the evolution mapping framework, the non-universality of the HMF can be framed in terms of its two main physical components. 
Following this approach, we model the multiplicity function by extending the universal ansatz to include an explicit dependence on the power spectrum shape, via $n_\mathrm{eff}$, and on the growth history, via $\tilde{x}$. 
Our full model for the multiplicity function is given by
\begin{equation}
    \label{eq:fnu_ours}
    f(\nu) = A_0\nu\Big( A\nu^a + B\nu^b\Big) \exp(-C\nu^2) \, , 
\end{equation}
where the dependence on $n_\mathrm{eff}$ and $\tilde{x}$ is encoded as
\begin{equation}
    \label{eq:fnu_ours_params}
    \begin{split}
    A &= 1 + A_n n_\mathrm{eff} \, , \\
    B &= B_0 + B_x \tilde{x} + B_n n_\mathrm{eff} \, , \\
    a &= a_n n_\mathrm{eff} \, , \\
    b &= b_n n_\mathrm{eff} \, .
    \end{split}
\end{equation}
There are thus nine parameters to be fitted to the HMF measured from simulations: eight of them appear in equations~(\ref{eq:fnu_ours}) and (\ref{eq:fnu_ours_params}), plus the memory parameter $\eta$, on which $\tilde{x}$ depends.
\begin{table*}
    \centering
    \caption{Cosmologies of the Aletheia simulations. The top block lists the parameters of the reference $\Lambda\mathrm{CDM}$ cosmology (Model 0). Models 1-8 are defined in the bottom block by varying one evolution parameter with respect to Model 0. Variations of $\omega_\mathrm{DE}$ also change the value of $h$ (by equation~(\ref{eq:hubble_sum}), while Model 7 has non-zero curvature, and the value of $\omega_\mathrm{DE}$ is adjusted consequently to match the reference value of $h$.}
    \begin{tabularx}{\textwidth}{YYYYYYYY}
    \hline
    \multicolumn{8}{c}{Reference $\Lambda\mathrm{CDM}$ parameters} \\
    $\omega_\mathrm{b}$ & $\omega_\mathrm{c}$ & $\omega_\mathrm{DE}$ & $\omega_K$ & $h$ & $n_\mathrm{s}$ & $w_0$ & $w_a$ \\
    \hline  
    0.02244 & 0.1206 & 0.3059 & 0 & 0.67 & 0.96 & -1 & 0 \\
    \hline
    \end{tabularx}
    \vspace{1em}
    \begin{tabularx}{\textwidth}{YW{0.3\textwidth}YYYYY}
    \hline
    \multirow{2}{*}{Model} & \multirow{2}{*}{Description and varied parameters} & \multicolumn{5}{c}{Redshifts of snapshots} \\
     & & $\sigma_{12}=0.343$ & $\sigma_{12}=0.499$ & $\sigma_{12}=0.611$ & $\sigma_{12}=0.703$ & $\sigma_{12}=0.825$ \\
    \hline
    Model 0 & Reference $\Lambda\mathrm{CDM}$ (top block) & 2.00 & 1.00 & 0.57 & 0.30 & 0.00 \\
    Model 1 & $\Lambda\mathrm{CDM}$, $\omega_\mathrm{DE} = 0.1594$ ($h=0.55$) & 1.76 & 0.86 & 0.48 & 0.25 & 0.00 \\
    Model 2 & $\Lambda\mathrm{CDM}$, $\omega_\mathrm{DE} = 0.4811$ ($h=0.79$) & 2.23 & 1.14 & 0.66 & 0.35 & 0.00 \\
    Model 3 & $w$CDM, $w_0 = -0.85$ & 2.10 & 1.04 & 0.59 & 0.31 & 0.00 \\
    Model 4 & $w$CDM, $w_0 = -1.15$ & 1.92 & 0.96 & 0.55 & 0.29 & 0.00 \\
    Model 5 & $w_0w_a$CDM, $w_a = -0.2$ & 1.97 & 0.99 & 0.57 & 0.30 & 0.00 \\
    Model 6 & $w_0w_a$CDM, $w_a = +0.2$ & 2.03 & 1.01 & 0.57 & 0.30 & 0.00 \\ 
    Model 7 & Non-flat $\Lambda\mathrm{CDM}$, $\Omega_K = -0.05$ ($\omega_\mathrm{DE} = 0.3283$) & 1.93 & 0.98 & 0.56 & 0.30 & 0.00 \\
    Model 8 & EdS model, $\omega_\mathrm{DE} = 0$ ($h=0.38$) & 1.40 & 0.65 & 0.35 & 0.17 & 0.00 \\
    \hline
    \end{tabularx}
    \label{tab:models_aletheia}
\end{table*}

\section{Measuring the halo mass function from cosmological simulations}
\label{sec:method}

\subsection{Simulations}
\label{subsec:sims}
We employ three suites of cosmological simulations. All are performed with \texttt{Gadget4} \citep{springel_gadget4_2021}, and employ paired and fixed initial conditions \citep{angulo_pontzen_paired_fixed_2016} to reduce the effect of cosmic variance on the simulated density field. The initial conditions were generated based on second-order Lagrangian perturbation theory with \texttt{2LPTic} \citep{crocce_2lptic_2012}.

\begin{itemize}
    \item The Aletheia suite consists of ten simulation pairs representing different cosmologies, all sharing the same set of shape parameters. 
    The different choices of evolution parameters are explained in detail in \citet{esposito_evolution_mapping_2024, sanchez_evo_mapping_2022}. 
    Model 0 represents our reference flat $\Lambda\mathrm{CDM}$ cosmology, with parameters fixed to the best-fitting values of \cite{planck_2018}, while the others (models 1 to 8) correspond to cosmological models strongly disfavoured by available data. 
    Their purpose is to demonstrate the validity of the evolution mapping approach well beyond the observationally viable ranges of the evolution parameters, as outlined in Table~\ref{tab:models_aletheia}.
    They contain $1500^3$ particles and have a box side length $L_{\rm{box}} = 1492.54$ Mpc, corresponding to $1000\,h^{-1} \mathrm{Mpc}$ for the reference value of $h = 0.67$. From each, we acquired five snapshots, at fixed values of $\sigma_{12}$. Having the linear power spectra of all models identical shape, fixing the value of $\sigma_{12}$ implies that at any given snapshot, the different models have \textit{exactly} the same linear matter power spectrum.
    Note that the same $\sigma_{12}$ can correspond to different redshift values in different cosmologies, as reported in Table~\ref{tab:models_aletheia}.
    
    \item The AletheiaMass simulations are our high-resolution $\Lambda\mathrm{CDM}$ suite. They represent the same cosmology as Model 0 of the Aletheia, with more particles and different box sizes. These include simulation pairs containing $2048^3$ particles, and $L_{\mathrm{box}} = \{180, \, 350, \, 700, \, 1400\}$ Mpc, and one with $2500^3$ particles and $L_{\mathrm{box}} = 2800$ Mpc. From each of these, we acquired ten snapshots at fixed values of $\sigma_{12}$, which are listed in Table~\ref{tab:redshifts_aletheiamass} together with their corresponding redshifts.
    
\begin{table}
    \centering
    \caption{Values of $\sigma_{12}$ and corresponding redshift, $z(\sigma_{12}, \vb{\Theta}_{\mathrm{s}},\vb{\Theta}_{\mathrm{e}})$, at each snapshot of the AletheiaMass simulations.}
    \begin{tabularx}{\columnwidth}{YY@{\hspace{0.15\columnwidth}}YY}
    \hline
    $\sigma_{12}$ & Redshift $z$ & $\sigma_{12}$ & Redshift $z$\\
    \hline
    0.2 & 4.20 & 0.7 & 0.30 \\
    0.3 & 2.46 & 0.8 & 0.06 \\
    0.4 & 1.56 & 0.825 & 0.00 \\
    0.5 & 1.00 & 0.9 & -0.17 \\
    0.6 & 0.61 & 1.0 & -0.39 \\
    \hline
    \end{tabularx}
    \label{tab:redshifts_aletheiamass}
\end{table}

    \item Finally, the AletheiaEmu represent a selection of $\Lambda\mathrm{CDM}$ models with different shape parameters and clustering amplitude. 
    The AletheiaEmu were performed to serve as training set for the \texttt{Aletheia} emulator of the non-linear matter power spectrum presented in \citet{sanchez_aletheia_2025}. 
    The full set consist of 150 nodes sampling the parameters $(\omega_{\mathrm{b}}, \omega_{\mathrm{c}},
    n_{\mathrm{s}}, \sigma_{12})$ on a Maximin Latin Hypercube design \citep{mckay_LH_1979,stein_MLH_1987}, with shape parameters covering a range of approximately $5\sigma$ from the best-fitting values of \citet{planck_2018}.
    The primordial scalar amplitude was kept fixed to $A_{\mathrm{s}} = 2.1\times 10^{-9}$, and the dimensionless Hubble parameter was fixed to $h =0.673$: consequently, $\omega_\mathrm{DE}$ varies to match the fixed value of $h$ given $\omega_{\mathrm{b}}$ and $\omega_{\mathrm{c}}$.
    We do not employ these simulations to calibrate our model. Rather, we will use a subset of them to test its accuracy in Section~\ref{subsec:external_datasets}.
    The cosmological parameters of the models we employ are summarised in Table~\ref{tab:models_aletheiaEmu}.
    
\begin{table}
    \centering
    \caption{Cosmologies of the AletheiaEmu simulations. Each model varies the shape parameters $\{ \omega_\mathrm{b}, \, \omega_\mathrm{c}, \, n_\mathrm{s} \}$, and the clustering amplitude $\sigma_{12}$ at which the snapshot is produced. The table also lists the corresponding redshift, $z(\sigma_{12}, \vb{\Theta}_{\mathrm{s}},\vb{\Theta}_{\mathrm{e}})$. Evolution parameters are common to all models and are listed in the bottom block, with the exception of $\omega_\mathrm{DE}$, which varies to match the common value of $h$.}
    \begin{tabularx}{\columnwidth}{YYYYYY}
        \hline
        Model & $\omega_\mathrm{b}$ & $\omega_\mathrm{c}$ & $n_\mathrm{s}$ & $\sigma_{12}$ & $z$ \\
        \hline
        C009 & 0.02218 & 0.1221 & 0.9483 & 0.272 & 2.81 \\
        C019 & 0.02185 & 0.1229 & 0.9556 & 0.352 & 1.94 \\
        C029 & 0.02197 & 0.1203 & 0.9652 & 0.426 & 1.38 \\
        C039 & 0.02219 & 0.1207 & 0.9821 & 0.510 & 0.96 \\
        C049 & 0.02160 & 0.1257 & 0.9477 & 0.593 & 0.67 \\
        C059 & 0.02240 & 0.1175 & 0.9617 & 0.670 & 0.35 \\
        C069 & 0.02286 & 0.1165 & 0.9875 & 0.749 & 0.14 \\
        C079 & 0.02243 & 0.1155 & 0.9878 & 0.831 & -0.06 \\
        C089 & 0.02249 & 0.1202 & 0.9526 & 0.916 & -0.23 \\
        C099 & 0.02226 & 0.1188 & 0.9584 & 0.995 & -0.42 \\
        \hline
        \hline
        \multicolumn{1}{c|}{\multirow{2}{0.16\columnwidth}{\makecell{Common \\ parameters}}} & $A_s \times 10^9$ & $w_0$ & $w_a$ & $\omega_K$ & $h$ \\
        \cline{2-6}
        \multicolumn{1}{c|}{}& $2.101$ & -1 & 0 & 0 & 0.673 \\
        \hline
    \end{tabularx}
    \label{tab:models_aletheiaEmu}
\end{table}
    
\end{itemize}

\subsection{Halo finding}
\label{subsec:rockstar}
We identify haloes in all our simulations with the \texttt{Rockstar} halo finder \citep{behroozi_rockstar_2013}. \texttt{Rockstar} starts by dividing the simulation box in particle groups performing a rough friends-of-friends \citep[FoF, ][]{fof_algorithm} procedure in position space. Then, it identifies groups of particles with a FoF iterative procedure in phase space, shortening the linking length at each step in order to identify progressively denser clumps, down to a minimal group size of 10 particles.
After such a hierarchy of phase-space densities has been built, its deepest levels are taken as halo seeds, and particles from higher levels are progressively assigned to the closest seed in phase space.

At this stage, \texttt{Rockstar} also performs a preliminary identification of subhaloes. For each halo, the closest larger phase-space neighbour in the original 3D FoF group is identified as the host, if any exist. The resulting substructure hierarchy is checked against the catalogues obtained at previous snapshots of the simulations, if available, to ensure that host and subhaloes do not get swapped from one snapshot to another, which would cause sudden extreme variations in their masses.  
The host-subhalo relationship is later confirmed once the halo masses are calculated, and only candidate subhaloes situated within the radius of their host are identified as subhaloes in the output catalogue.
This allows to build a merger tree, identifying subhaloes.

We make use of this information to clean our halo catalogues in post-processing, excluding all the subhaloes that have their centre of mass within the Lagrangian radius of their parent, i.e., if
$| \vb{x}_{\rm{sub}} -  \vb{x}_{\rm{parent}} | < R_\mathrm{parent}$. Note that this means that haloes are allowed to partially overlap, although no halo contains the centre of mass of another one. This criterion is the same as used by \citet{tinker_hmf_2008}. 
The fraction of haloes excluded by this procedure from the halo catalogues of the simulations used in this work lies between 5 and 10 per cent, with a general increasing trend of the excluded subhalo fraction over time.

Before computing halo masses by growing spheres from the identified centres up to the desired overdensity threshold, \texttt{Rockstar} removes from each halo all particles which are not gravitationally bound to it. 
As a consequence, there are particles within the radius of a halo whose mass is not counted in it. 
The code allows this option to be disabled, returning the total mass enclosed within the spherical overdensity. 
The resulting multiplicity functions differ by 2–3 per cent between the two cases.

An all-inclusive spherical overdensity definition may be more appropriate to compare the HMF with observed cluster counts. 
However, for the purpose of characterising the response of the mass function to differences in the history of structure formation and to the local shape of the linear power spectrum, a bound-only halo definition is preferable, as it avoids the effects of transient or recently stripped material on the measured abundances. 

Therefore, throughout the main body of this paper we show results obtained excluding unbound particles from haloes, i.e., setting \texttt{STRICT\_SO\_MASSES~=~0} in the \texttt{Rockstar} configuration parameters file. 
The calibration of the HMF with haloes including all particles within the overdense sphere is presented in Appendix~\ref{appendix_strict_SO}.

\subsection{Binning and estimating the multiplicity function}
\label{subsec:binning}

\begin{figure}
    \centering
    \includegraphics[width=0.95\linewidth]{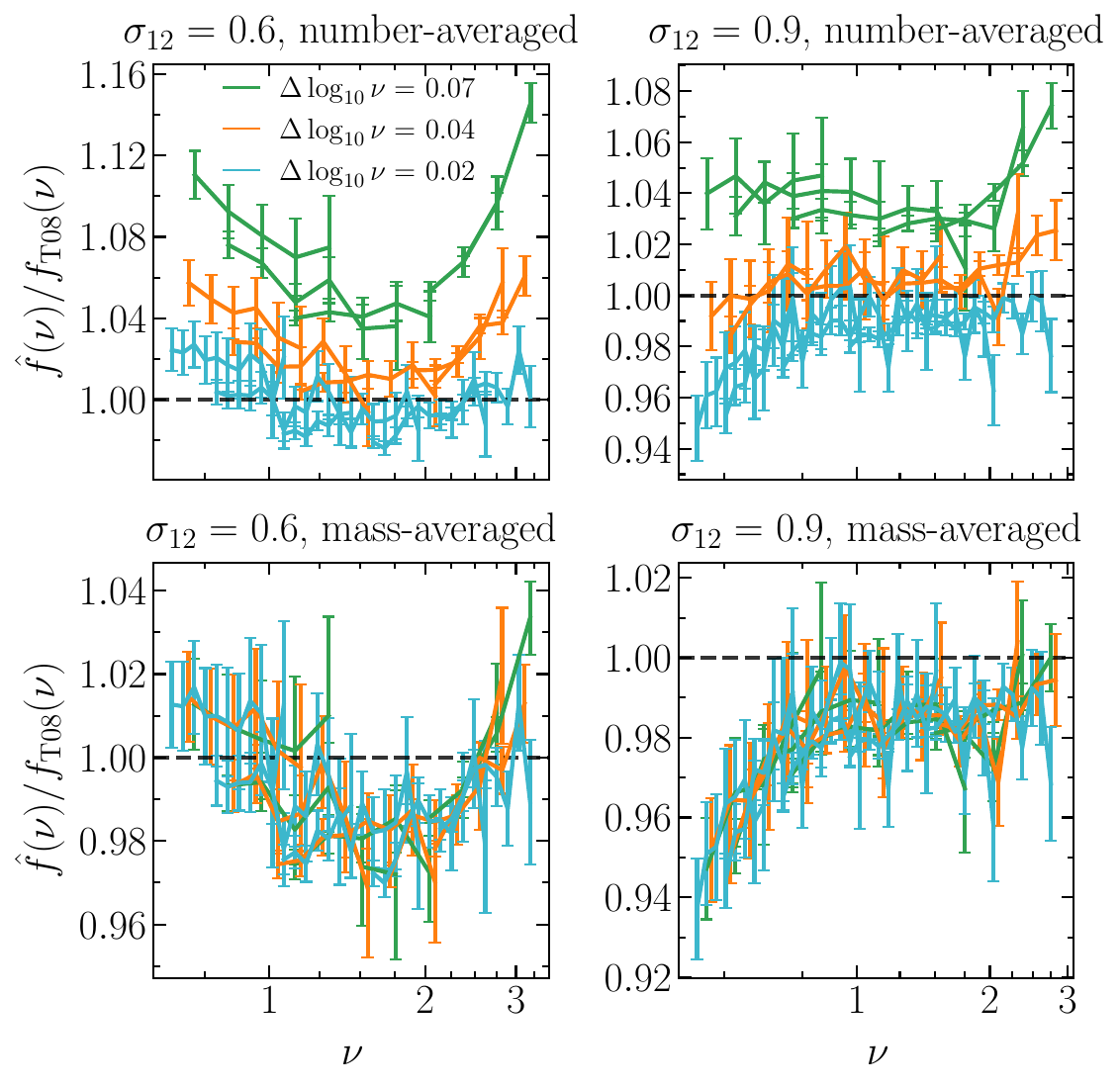}
    \caption{Stability of the measurements of the multiplicity function with respect to changes in the binning scheme, in two example snapshots of the AletheiaMass simulations. Assigning to each bin a representative mass value $M(\nu)$ evaluated at the central $\nu$ of the bin, as shown in the upper panels, can bias the estimate by a few per cent. Instead, summing masses directly into the estimator, as in equation~(\ref{eq:fnu_estimator_masses}), does not, as the lower panels show.}
    \label{fig:binning_stability}
\end{figure}

Binning the halo catalogue to measure $f(\nu)$ can introduce a bias in the measurement, which becomes particularly evident when the binning is coarse \citep{Li_Smith_2025}. 
We demonstrate this in the upper panels of Fig.~\ref{fig:binning_stability}, where we show the multiplicity function from two snapshots of the AletheiaMass simulations with different bin widths.
We plot the measurements as a ratio to the fitting function of \citet{tinker_hmf_2008} to reduce the dynamic range of the $y$ axis and make the difference between the lines more apparent. The measured values increase systematically with bin width.

This bias originates from how $ f(\nu) $ is estimated. 
Counting the number of haloes in one mass bin and dividing by the volume of the simulation and the bin width gives an unbiased estimate of the integral average of $dn/dM$ over the bin. 
Since $dn/dM$ is a smooth function of $M$, the mean value theorem ensures the existence of a mass $ M_* $ within the bin for which the function value equals the bin average. 
However, this $ M_* $ does not, in general, coincide with the central mass of the bin, $ M_\mathrm{c} $, or with any other representative mass one might choose for the bin, such as the mass corresponding to the central peak height, $M(\nu_\mathrm{c})$, and this mismatch introduces a bias.

When transforming the halo counts into an estimate of $f(\nu)$ using equation~(\ref{eq:fnu_dndm}), one typically multiplies the count-derived estimate by a representative mass for the bin, most commonly $M_\mathrm{c}$. 
The mismatch between $M_\mathrm{c}$ and $M_*$ biases the product, i.e., the estimate of $f(\nu)$.

This can be prevented using the average mass of haloes within the bin as representative value, i.e., directly estimating the integral of the mass density, $M \, dn/dM$, rather than of the number density. 
This translates into an estimator of the multiplicity function of the form:
\begin{equation}
    \label{eq:fnu_estimator_masses}
    \hat{f}(\nu) = \frac{1}{\overline{\rho}\,  V_\mathrm{box}}\frac{1}{\Delta \ln \nu}\sum_{\nu_i \in [\nu_\mathrm{l}, \nu_\mathrm{r})} M_i \, . 
\end{equation}
where $V_{\rm box}={L_{\rm box}}^3$ is the volume of the simulated box.
The bottom panel of Fig. \ref{fig:binning_stability}
shows how the stability of the estimates with respect to the binning choice improves when adopting the estimator of equation~(\ref{eq:fnu_estimator_masses}), for which no appreciable differences arise between measurements adopting different binning schemes.

When referring to the multiplicity function measured from simulations, we will adopt the notation $\hat{f}(\nu)$, thereby implying it was measured using equation~(\ref{eq:fnu_estimator_masses}) as an estimator. 
We leave the notation $f(\nu)$ for the theoretical multiplicity function defined by equation~(\ref{eq:fnu_dndm}), and in particular we will refer to numerically calibrated models (from this or other works) as $f_\mathrm{fit}(\nu)$.

To keep the scatter of our estimates and their Poisson errors under control, we bin the multiplicity function using 10 bins of equal logarithmic width (i.e., equally spaced in $\log \nu$) in the range $\nu \in [0.4, 1)$, and 16 linearly spaced bins in the range $\nu \in [1, 4]$. We assign to every bin a representative value given by its log-centre. Note that the same $\nu$ bin corresponds to different mass ranges at different $\sigma_{12}$.

\begin{figure}
    \centering
    \includegraphics[width=\linewidth]{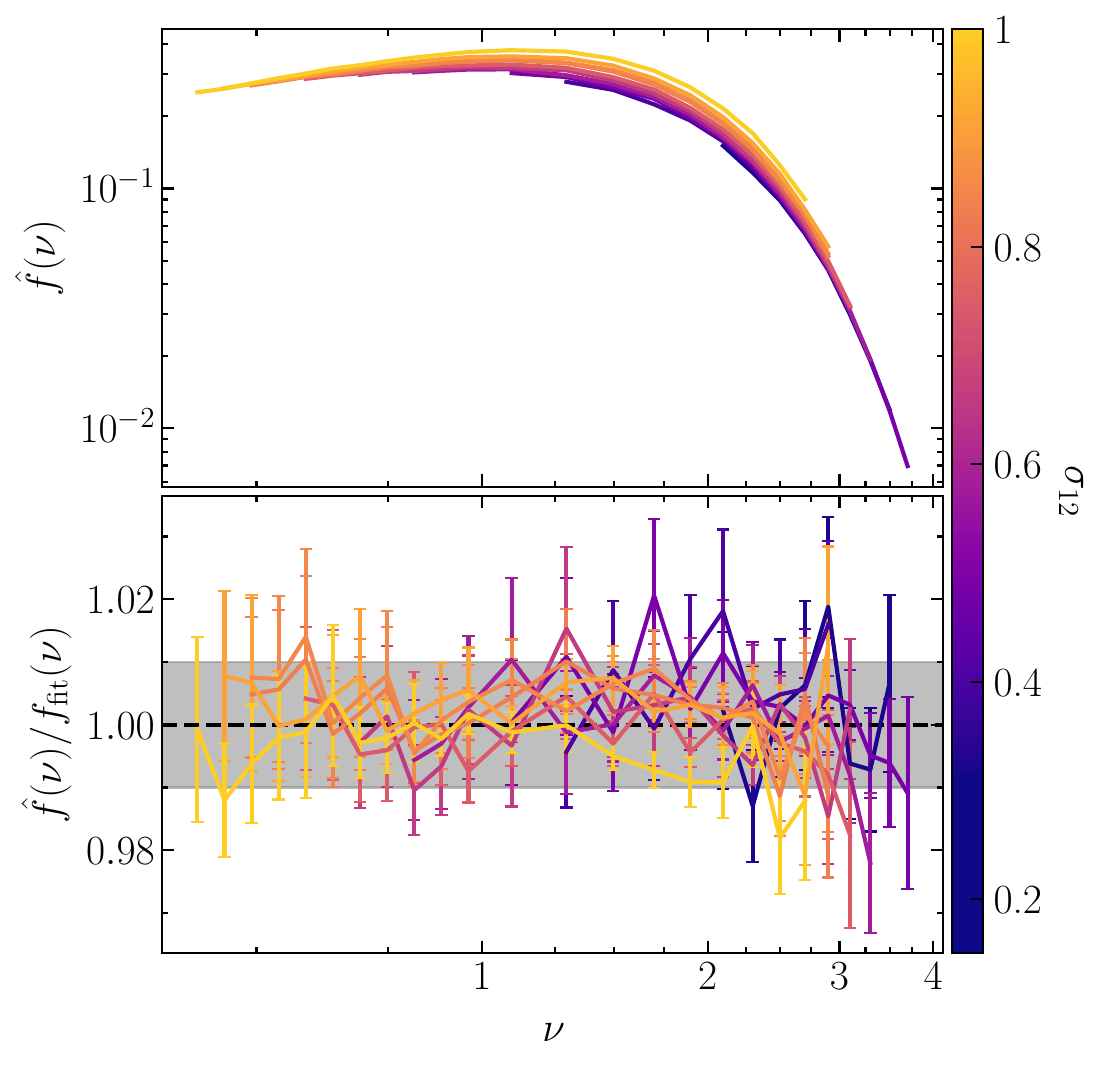}
    \caption{Upper panel: multiplicity function $\hat{f}(\nu)$ measured in the AletheiaMass simulations, corresponding to an overdensity threshold of $\Delta = 200$. Lower panel: ratios of the measured values to the calibrated fitting function given in equation (\ref{eq:fnu_ours}). 
    To avoid crowding the plots with the measurements from all the AletheiaMass simulations, here and in subsequent figures we show, at each snapshot and for each mass bin, a weighted average of the measurements from different boxes. 
    The grey band in the lower panel corresponds to a 1 per cent difference.}
    \label{fig:residuals_aletheiamass_m200b}
\end{figure}

\subsection{Uncertainties}
\label{subsec:uncertainties}
To quantify the uncertainty on our estimates $\hat{f}(\nu)$ we take into account both the Poisson noise related to halo counts and the sample variance introduced by the finite box size \citep{hu_kravtsov_sample_variance_2003, crocce_hmf_2010, smith_marian_counts_covariance_2011}. 
The relative error on the HMF in a bin centred around $\nu$ can be written as: 
\begin{equation}
    \label{eq:fnu_relative_variance}
    \frac{\sigma_{\hat{f}(\nu)}}{\hat{f}(\nu)} = {\bigg[ b^2(\nu) \sigma^2(R_\mathrm{eff}) + \frac{1}{N_\mathrm{haloes}(\nu)}\bigg]}^{1/2} \, .
\end{equation}
Here, $\sigma^2(R_\mathrm{eff})$ is the RMS density fluctuation on the scale of the simulated box, the effective radius being that of a sphere with the same volume as the simulation, $ R_\mathrm{eff} = {(3V_\mathrm{box}/4\pi)}^{1/3} $. 
The function $b(\nu)$ represents the linear halo bias, which we describe
using the prescription of \citet{tinker_bias_2010}.

Although the impact of cosmic variance on statistics of the matter density field is largely suppressed when running simulations with paired and fixed initial conditions, \citet{villaescusa_paired_fixed_2018} showed that the impact on the HMF variance is negligible at high masses, $M \gtrsim 10^{13} M_\odot$.
At lower masses, fixing the amplitude of the power spectrum when setting the initial conditions can reduce $\sigma_{\hat{f}(\nu)}$ by up to 50 per cent.
When fitting the parameters of our prescription, equation~(\ref{eq:fnu_ours}), we set to zero the off-diagonal terms of the covariance of $\hat{f}(\nu)$.

\begin{figure}
    \centering
    \includegraphics[width=0.95\linewidth]{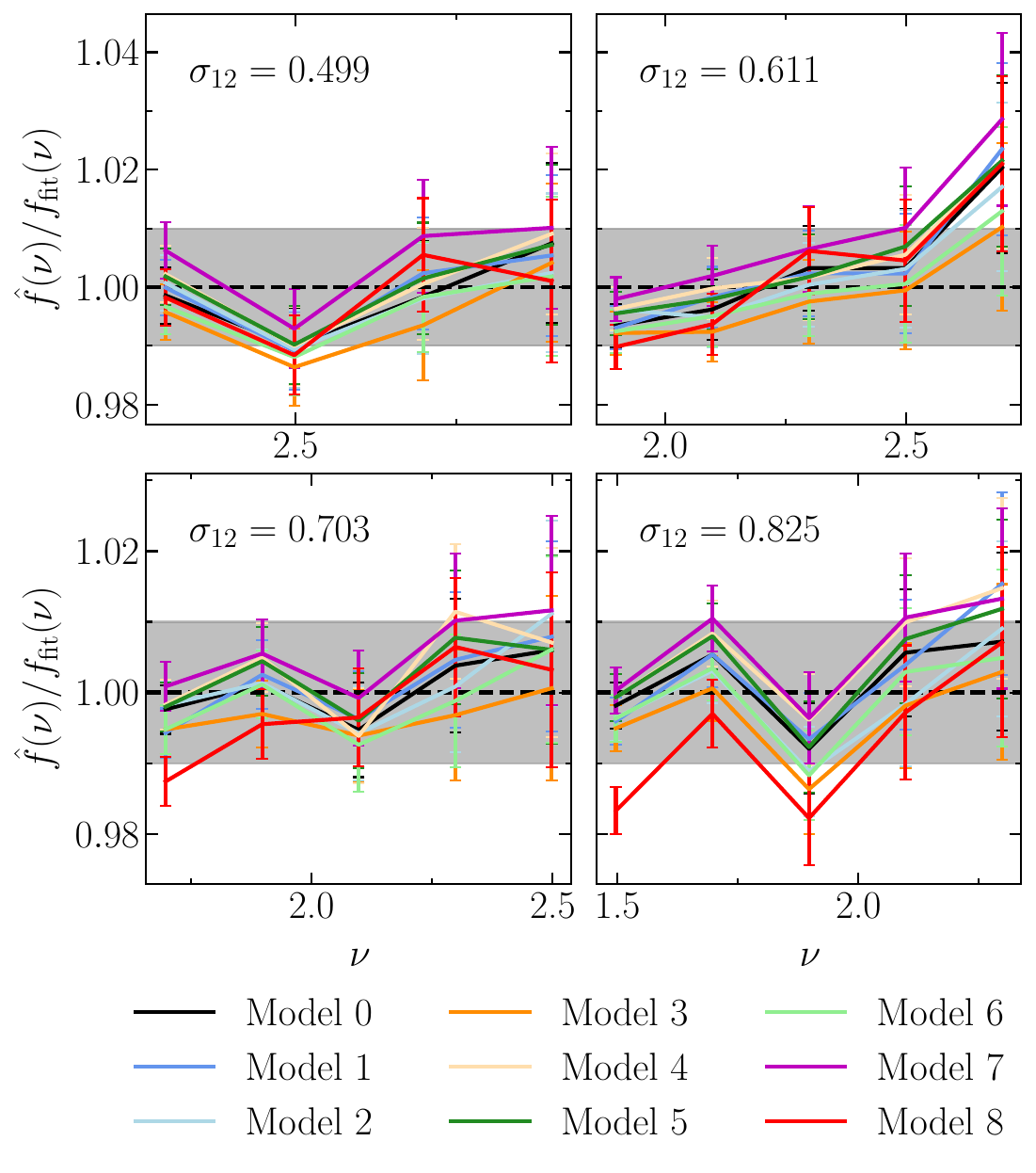}
    \caption{Ratios of $\hat{f}(\nu)$ measured in the Aletheia simulations for an overdensity threshold $\Delta = 200$ and  the calibrated fitting function given by equation~(\ref{eq:fnu_ours}).}
    \label{fig:residuals_aletheia_m200b}
\end{figure}

\subsection{Convergence of the measurements}
\label{subsec:convergence}
The low-mass tail of the measured HMF is subject to systematic biases due to the finite mass resolution of the simulations. Haloes composed of only a small number of particles are either not robustly identified by halo-finding algorithms or are absent altogether, leading to an artificial suppression of $\hat{f}(\nu)$ at low $\nu$ \citep[e.g.,][]{jenkins_hmf_2001, angulo_extending_halo_mass_resolution_2014}.
This is particularly problematic when calibrating semi-analytic models against simulation data, since the resulting bias is not reflected in the statistical uncertainties, and can therefore propagate into a miscalibrated fitting function. 

To prevent this, it is common practice to impose a minimum number of particles per halo, $N_\mathrm{min}$, below which haloes are excluded from the catalogues \citep[e.g., ][]{jenkins_hmf_2001}. For haloes identified with \texttt{Rockstar}, $N_\mathrm{min}$ lies in the order of 100 \citep[see ][for a systematic test using scale-free simulations and the virial mass]{maleubre_haloes_scale_free_2024}. On the other hand, $N_\mathrm{min}$ depends on the overdensity threshold $\Delta$ chosen to define the halo boundary \citep{tinker_hmf_2008}.

In our analysis, the AletheiaMass suite provides a useful testing ground for assessing convergence: by comparing simulations of varying box sizes, we can empirically determine the scale at which the HMF remains stable under changes in resolution. We adopt a deliberately conservative choice of $N_\mathrm{min}$ in the smallest box and enforce consistency with larger-volume runs, within the uncertainties given by equation~(\ref{eq:fnu_relative_variance}). For the Aletheia and AletheiaEmu simulations, which lack higher-resolution counterparts, we employ conservative thresholds to ensure robust measurements across the full halo mass range considered. For this reason, in many of the figures showing measurements from the Aletheia suite, the first snapshot is absent. At such low clustering amplitude, we do not find any bins with haloes containing more particle than $N_\mathrm{min}$ and with enough haloes for the Poisson error to remain reasonably small, given the resolution of the Aletheia simulations.
The required values of $N_\mathrm{min}$ are listed in Table~\ref{tab:nmin_haloes}.

\begin{figure*}
    \centering
    \includegraphics[width=0.9\textwidth]{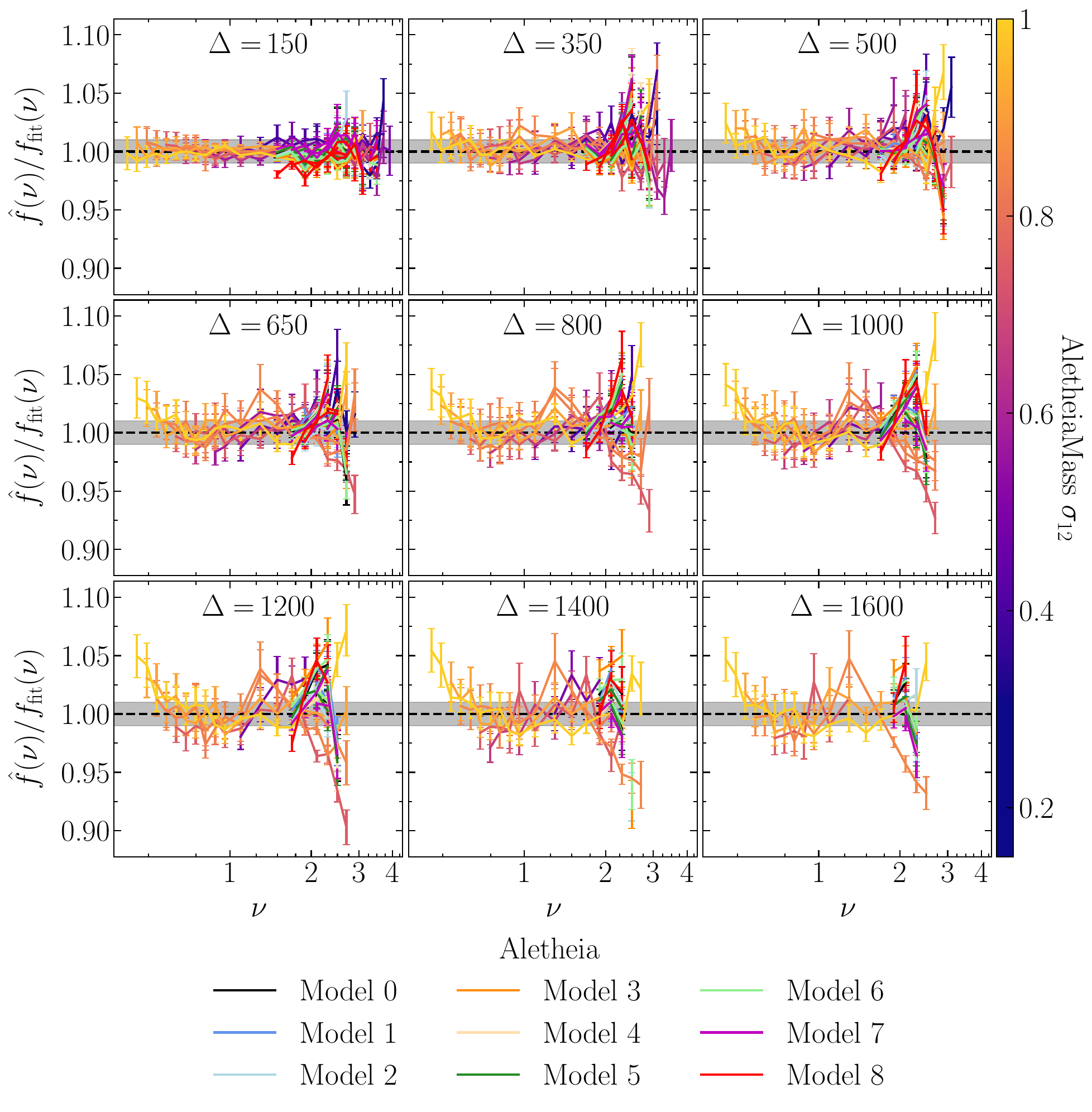}
    \caption{Ratios of $\hat{f}(\nu)$ measured in the Aletheia and AletheiaMass simulations to the calibrated fitting function given in equation~(\ref{eq:fnu_ours}), for nine different values of the overdensity threshold $\Delta$. 
    The grey bands correspond to a 1 per cent difference.
    }
    \label{fig:fit_allmasses}
\end{figure*}

\begin{table}
    \centering
    \caption{Minimum required number of particles per halo  as a function of the overdensity threshold $\Delta$.}
    \begin{tabularx}{\columnwidth}{Y|YYYYYYYYYY}
        \hline
        $\Delta$ & 150 & 200 & 350 & 500 & 650 & 800 & 1000 & 1200 & 1400 & 1600  \\
        \hline
        $N_\mathrm{min}$ & 1000 & 1000 & 2000 & 2500 & 2500 & 3000 & 3000 & 3000 & 3500 & 4000  \\
        \hline
    \end{tabularx}
    \label{tab:nmin_haloes}
\end{table}

\section{Calibration of the model}
\label{sec:results}

We begin by showing an example of the resulting $\hat{f}(\nu)$ from the method described in Section~\ref{sec:method} in the upper panel of Fig.~\ref{fig:residuals_aletheiamass_m200b}.
The measurements from the AletheiaMass simulations, with haloes defined by $\Delta = 200$, give an illustration of the non-universality of the HMF, which shows a clear time dependence.

\begin{table*}
    \centering
    \caption{Best-fitting parameters of our model multiplicity function, given in equation~(\ref{eq:fnu_ours}), for for the different overdensity thresholds $\Delta$ considered in this work.}
    \begin{tabularx}{\textwidth}{Y|YYYYYYYYY}
        \hline
         $\Delta$ & $A_0$ & $A_n$ & $a_n$ & $B_0$ & $B_x$ & $B_n$ & $b_n$ & $C$ & $\log_{10} \eta$ \\
         \hline
         150 & 0.42 & 0.040 & 0.232 & -0.6 & 1.21 & 0.148 & -0.59 & 0.431 & -0.37  \\
         200 & 0.47 & 0.082 & 0.234 & -0.6 & 1.03 & 0.102 & -0.63 & 0.445 & -0.54  \\
         350 & 0.62 & 0.164 & 0.219 & -0.6 & 0.77 & 0.035 & -0.76 & 0.487 & -0.74  \\
         500 & 0.68 & 0.200 & 0.22 & -0.6 & 0.68 & 0.009 & -0.8 & 0.517 & -0.92  \\
         650 & 0.71 & 0.221 & 0.228 & -0.6 & 0.64 & -0.003 & -0.82 & 0.543 & -1.08  \\
         800 & 0.79 & 0.248 & 0.22 & -0.6 & 0.57 & -0.026 & -0.86 & 0.561 & -1.13  \\
         1000 & 0.86 & 0.272 & 0.212 & -0.6 & 0.51 & -0.049 & -0.83 & 0.576 & -1.40  \\
         1200 & 0.85 & 0.273 & 0.215 & -0.6 & 0.51 & -0.047 & -0.84 & 0.599 & -1.65  \\
         1400 & 0.87 & 0.286 & 0.215 & -0.6 & 0.46 & -0.064 & -0.79 & 0.60 & $< - 2.0$ \\
         1600 & 0.89 & 0.294 & 0.23 & -0.6 & 0.47 & -0.06 & -0.83 & 0.63 & $< - 2.4$ \\
         \hline
    \end{tabularx}
    \label{tab:fitting_parameters}
\end{table*}

We calibrate the HMF model of equation~(\ref{eq:fnu_ours}), introduced and motivated in Section~\ref{sec:modelling}, using $\hat{f}(\nu)$ measured in the AletheiaMass and Aletheia simulations, for multiple values of the overdensity threshold $\Delta$, and provide a recipe to interpolate across the mass definitions, so that the model $f_\mathrm{fit}(\nu)$ can be computed at any value of $\Delta \in [150, 1600]$.

\begin{figure*}
    \centering
    \includegraphics[width=0.9\textwidth]{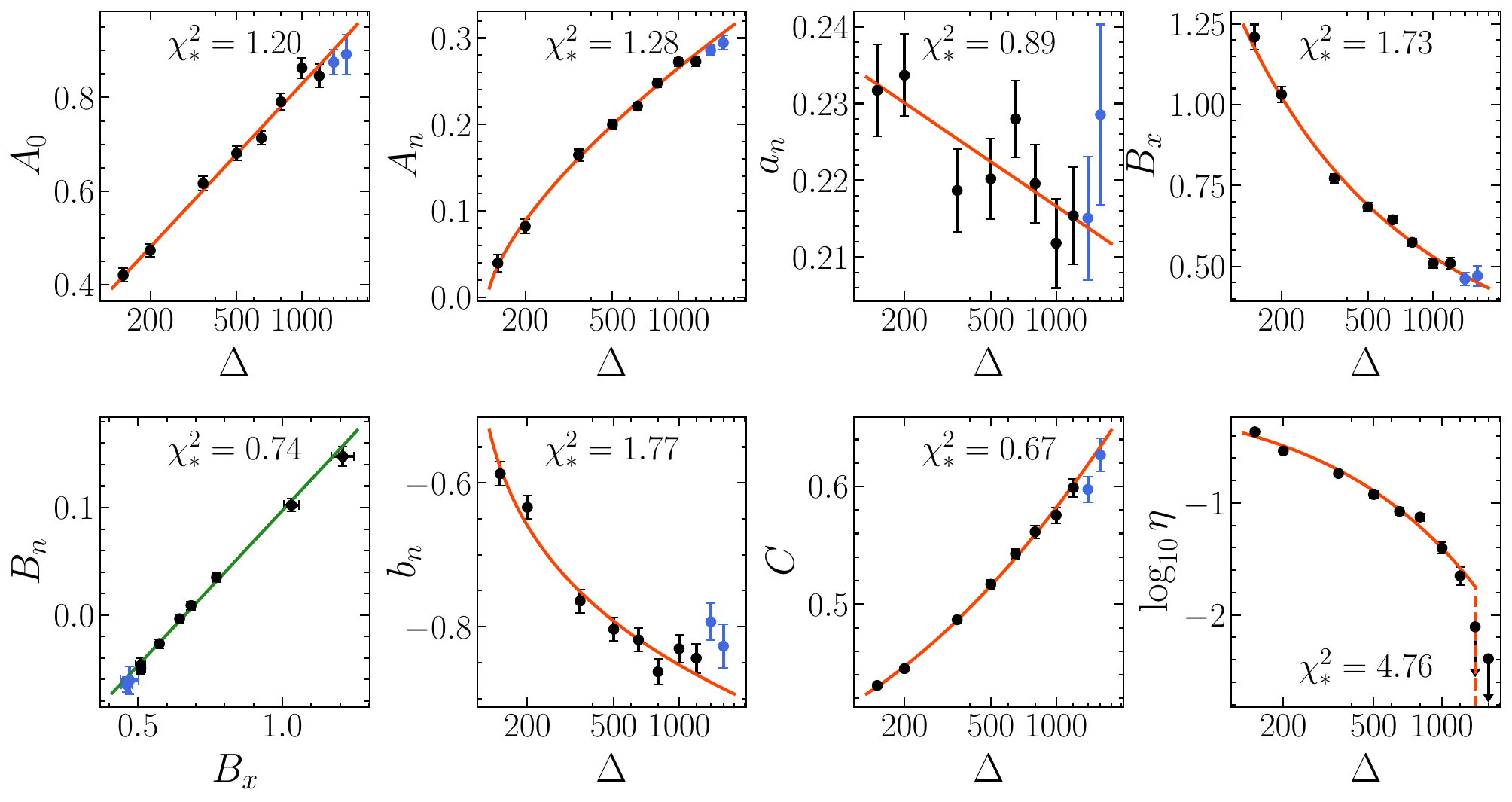}
    \caption{Interpolation of the best-fitting parameters as a function of the overdensity threshold. The black points represent the values listed in Table~\ref{tab:fitting_parameters}, while lines represent the empirical interpolating functions given in equation~(\ref{eq:interpolation_fitting_parameters}). The fitting function for the memory parameter $\eta$ is truncated at $\Delta = 1400$, above which we set $\eta = 0$ (see Section~\ref{subsec:fit_all}). For all other parameters, we exclude the values obtained from the two highest overdensity thresholds ($\Delta = 1400, \, 1600$, blue points) and use them to test the extrapolation to high overdensities. Interpolations as a function of
    $y=\log_{10}\Delta$ are shown as red lines, while $B_n$ interpolated a function of $B_x$ is shown in green. In each panel, we report the reduced chi-square of the respective fit, $\chi^2_* = \chi^2 / \mathrm{d.o.f.}$.}
    \label{fig:interpolation_fitting_parameters}
\end{figure*}

We derive posterior distributions for the free parameters using the importance nested sampling package \texttt{Nautilus} \citep{lange_nautilus_2023}.
We require a minimum effective number of points of $10^4$.
To avoid any biasing of the posterior distributions due to the samples collected during the exploration phase, we discard them and only keep points collected in the sampling phase.
We checked that neither the parameter constraints nor the Bayesian evidence change if the minimum effective sample size is increased to $10^6$.

As the mass bins we use for the calibration always contain $\gtrsim 1000$ haloes, we assume the likelihood of $\hat{f}(\nu)$ to be Gaussian, with the diagonal covariance introduced in Section~\ref{subsec:uncertainties}. 
We set wide flat priors on all parameters, except for the memory parameter $\eta$: as we do not have any prior knowledge about its scale, we set a log-uniform prior on its value.

\subsection{Fit for $M_{200\mathrm{b}}$}
\label{subsec:fit_m200b}
We first present the calibration of $f(\nu)$ for our baseline halo catalogues, defined with $\Delta = 200$.
The ratios of the multiplicity functions measured in the AletheiaMass simulations to the fit are shown in the lower panel of Fig.~\ref{fig:residuals_aletheiamass_m200b}.
In this and all other figures showing $\hat{f}(\nu)$ from the AletheiaMass, at each snapshot we show the variance-weighted average of measurements from different boxes in overlapping mass bins, following equation~(\ref{eq:fnu_relative_variance}), to avoid overcrowding the plots.

The model reaches per cent accuracy on our $\Lambda\mathrm{CDM}$ simulations, across the whole range of masses and at all redshifts, with very few points falling out of the 1 per cent grey band.

Turning to cosmologies with different structure formation histories, Fig.~\ref{fig:residuals_aletheia_m200b} shows the ratios $\hat{f}(\nu) / f_\mathrm{fit}(\nu)$ of measurements from the Aletheia simulations.
Again, residuals of all models in the four snapshots rarely exceed 1 per cent, the maximum shift from the model across all data points used for its calibration being of 3 per cent. 
Importantly, the only difference in the parameters we use to model the HMF between two different Aletheia cosmologies at the same $\sigma_{12}$ lies in $\tilde{x}$, as the linear power spectra are the same, and so are the values of $n_\mathrm{eff}$ and the mapping of the $\nu$-bins to mass space.
The ratios shown in Fig.~\ref{fig:residuals_aletheia_m200b}, compared to the ratios of measurements to Model 0 of Fig.~\ref{fig:aletheia_ratio_tomodel0}, show the importance of taking into account structure formation histories in the HMF: the spread across models, reaching 5 per cent already at early times and 10 per cent in later snapshots in Fig.~\ref{fig:aletheia_ratio_tomodel0}, is reduced to sub-per cent discrepancies, consistent within the error bars of the measurements, when including $\tilde{x}$ in the model.

\subsection{Fit for other mass definitions and parameters interpolation}
\label{subsec:fit_all}

We refit the model parameters on the measurements obtained from halo catalogues defined with different overdensity thresholds, ranging from $\Delta = 150$ to $\Delta = 1600$.

We report the resulting values in Table~\ref{tab:fitting_parameters}, while in Fig.~\ref{fig:fit_allmasses} we show the ratios of the measurements from the Aletheia and AletheiaMass simulations to the fitted multiplicity function for all measured values of $\Delta$. 
At high $\Delta$, the HMF is not well converged in the simulations, and the criteria introduced in Section~\ref{subsec:convergence} leave us with few data points: constraining the parameters thus becomes increasingly difficult. 
In particular, when fitting $\hat{f}(\nu)$ of haloes defined with the two highest values of $\Delta$, we only find an upper limit for the memory parameter $\eta$.

The fits retain a very high accuracy across the whole range of mass definitions, exceeding a relative shift of 5 per cent in only a handful of bins at high $\Delta$.

To make the HMF model applicable at arbitrary overdensity thresholds, we provide empirical interpolation formulae for the best-fitting parameters as continuous functions of $\Delta$.
Specifically, we express their scaling with $y = \log_{10}\Delta$ as:

\begin{equation}
\begin{split}
        &A_0 = 0.50 y - 0.67 \, , \\
        &A_n = 0.292 \, {(y - 2.13)}^{0.67} \, , \\
        &a_n = - 0.019 y + 0.27 \, , \\
        &B_x = 8.1 y^{-2.48} \, , \\
        &B_n = 0.29 B_x - 0.19 \, , \\
        &b_n = -0.86 \, {(y - 2.0)}^{0.20} \, , \\
        &C = 0.07 y^2 - 0.17 y + 0.5 \, , \\
        &\log_{10} \eta = -1.40 \, {\Big(\frac{\Delta}{1000}\Big)}^{0.65} \, . 
    \label{eq:interpolation_fitting_parameters}
\end{split}
\end{equation}

The coefficients were obtained by minimizing the $\chi^2$ on the fitted parameter values across the calibrated mass definitions.
When computing the model for one of the values of $\Delta$ listed in Table \ref{tab:fitting_parameters}, we recommend using the values obtained from the fit rather than the interpolated ones.

\begin{figure}
    \centering
    \includegraphics[width=0.95\linewidth]{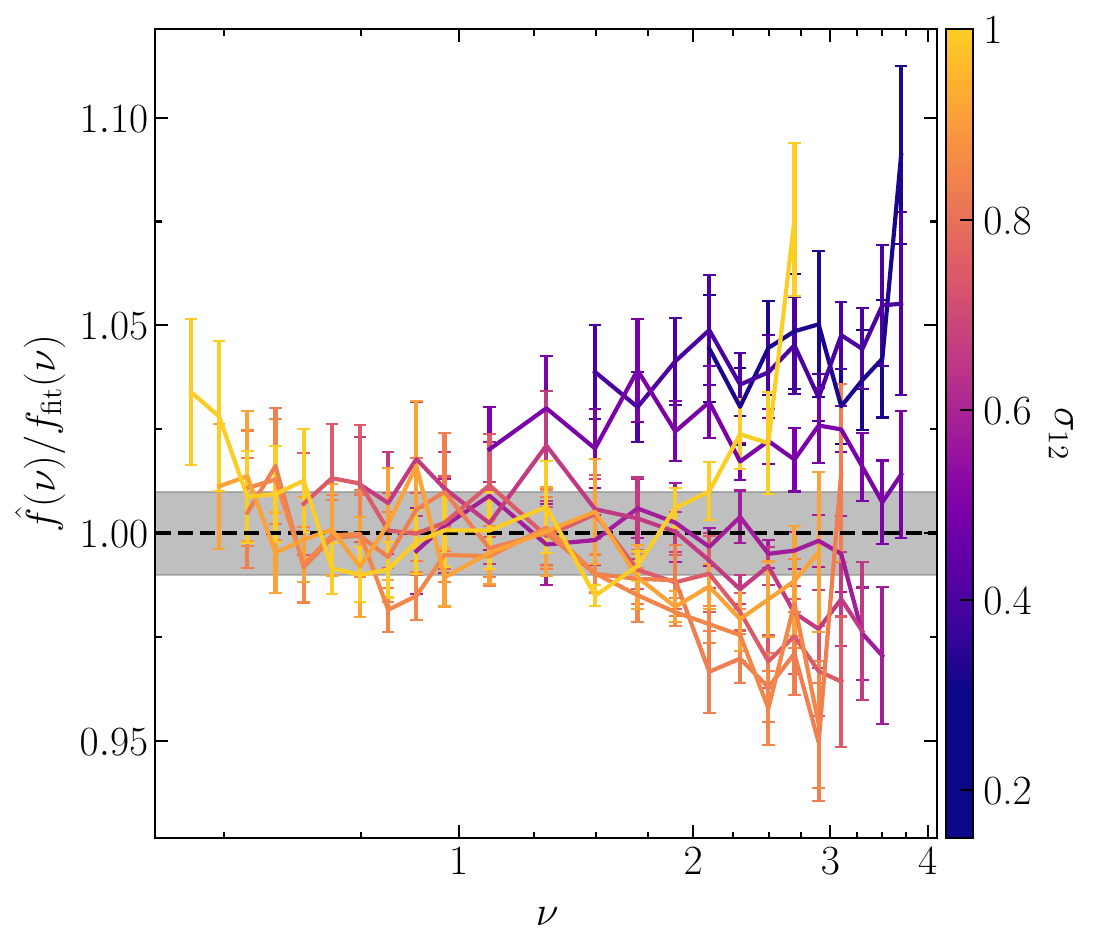}
    \caption{Ratios of the $\hat{f}(\nu)$ defined in terms of the virial mass measured in the AletheiaMass simulations to the fitting function given in equation~(\ref{eq:fnu_ours}), using parameter values interpolated to the corresponding value of $\Delta_\mathrm{vir}$ according to equation~(\ref{eq:interpolation_fitting_parameters}). The grey band corresponds to a 1 per cent difference.}
    \label{fig:residuals_aletheiamass_vir}
\end{figure}

As previously mentioned, at $\Delta = 1400,~1600$ we can only obtain upper bounds for the value of $\eta$: the constraints we get are, respectively, $\eta < 10^{-2.0}$ and $\eta < 10^{-2.4}$ within $95$ per cent posterior probability.
In practice, a low value of $\eta$ makes the Gaussian kernel of equation~(\ref{eq:x_tilde_def}) sharply peaked around the instantaneous value of $x(z)$. 
The relative difference between $x$ and $\tilde{x}$ is already well below 1 per cent when $\eta = 10^{-2}$, in our reference $\Lambda\mathrm{CDM}$ cosmology at $z=0$. 
We therefore choose to set $\eta = 0$ when computing the model for $\Delta \ge 1400$, i.e., we substitute $\tilde{x}$ with $x(z)$ at such high overdensities.

\begin{figure}
    \centering
    \includegraphics[width=0.95\linewidth]{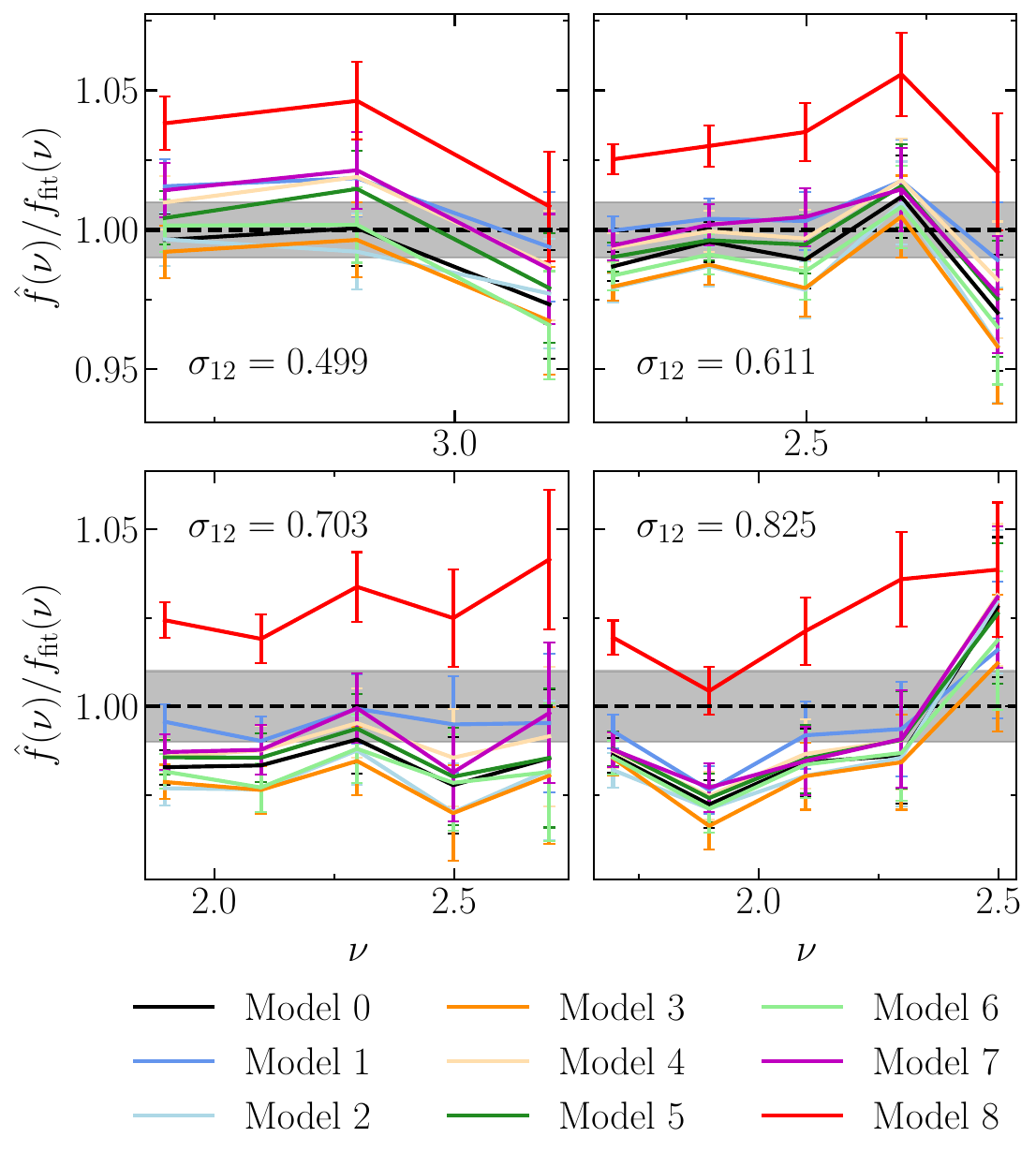}
    \caption{Ratios of the $\hat{f}(\nu)$ defined in terms of the virial mass measured in the Aletheia simulations to the fitting function given in equation~(\ref{eq:fnu_ours}), using parameteter values interpolated to the corresponding value of  $\Delta_\mathrm{vir}$ according to equation~(\ref{eq:interpolation_fitting_parameters}). The grey bands correspond to a 1 per cent difference.}
    \label{fig:residuals_aletheia_vir}
\end{figure}

In Fig.~\ref{fig:interpolation_fitting_parameters}, we show the accuracy of equation~(\ref{eq:interpolation_fitting_parameters}) interpolating the parameters' fitted values. 
The two highest overdensities, $\Delta = 1400,~1600$, were excluded from the calibration of the interpolating functions, and from the reduced chi-square values quoted in every panel, and used to check how well the scaling relations can be extrapolated above the probed range of thresholds: as the blue points in Fig.~\ref{fig:interpolation_fitting_parameters} show, the empirical scaling reproduces the trends remarkably well up to $\Delta = 1600$. 
We advise caution in extrapolating the parameters to much higher values of $\Delta$.

Computing the model with interpolated parameters still provides a very accurate fit to the simulations. We demonstrate this by testing our model on the virial halo catalogues obtained from the Aletheia and AletheiaMass simulations. We employ the virial overdensity threshold fit given by \citet{bryan_norman_1998}:
\begin{equation}
    \Delta_\mathrm{vir} = 18 \pi^2 + 82\, \big(\Omega_\mathrm{m}(z) - 1\big) - 39\, {\big(\Omega_\mathrm{m}(z) - 1\big)}^2 \, .
    \label{eq:virial_delta}
\end{equation}
As the threshold $\Delta_\mathrm{vir}$ of equation~(\ref{eq:virial_delta}) is cosmology- and time-dependent\footnote{Strictly speaking, equation~(\ref{eq:virial_delta}) is only valid for flat universes. For the purpose of demonstrating how well the HMF model holds, we also adopt it for Model 7 of the Aletheia simulations, which has non-zero curvature.}, it provides an excellent testing ground for our interpolation of the parameter values.

Figure~\ref{fig:residuals_aletheiamass_vir} shows the ratios of the measurements from the AletheiaMass simulations relative to our interpolated fitting function (equation~(\ref{eq:fnu_ours}) with the parameters computed from equation~(\ref{eq:interpolation_fitting_parameters})). 
The accuracy is nearly always within 5 per cent, but one can notice a slight trend in the ratios with increasing $\sigma_{12}$, excluding the last two snapshots. 
Figure~\ref{fig:residuals_aletheia_vir} shows the ratios for the Aletheia simulations over the same interpolated fit. 
While the accuracy is also approximately 5 per cent, the models appear in a reversed order compared to the original non-universality pattern seen in Fig.~\ref{fig:aletheia_ratio_tomodel0}. 
This reversal strongly suggests that our parameter interpolation is not perfectly capturing the dependence of the HMF on the growth history parameter, $\tilde{x}$. 
This limitation likely explains the slight trend in Fig.~\ref{fig:residuals_aletheiamass_vir}. 
The interpolation of our fitted parameters (which was calibrated on other mass definitions) does not capture the dependence of the virial HMF 
on $\tilde{x}$ with the same precision as a direct fit.

\section{Discussion}
\label{sec:discussion}

\subsection{Validation on additional datasets}
\label{subsec:external_datasets}

In this Section, we test the accuracy of our new prescription for $f(\nu)$ against measurements from simulations we have not used for its calibration.

The AletheiaEmu simulations, introduced in Section~\ref{subsec:sims} were designed to train the \texttt{Aletheia} emulator of the non-linear matter power spectrum presented in \citet{sanchez_aletheia_2025}, to which we refer for a more detailed description.
Exploiting this full suite to build an emulator of the HMF is one of the directions of our future work. 
In this paper, we use a subset of the AletheiaEmu simulations, summarised in Table~\ref{tab:models_aletheiaEmu}, to check the performance of our prescription  $f_\mathrm{fit}(\nu)$ on different cosmologies, varying both shape and evolution parameters from the ones it was calibrated on.

\begin{figure}
    \centering
    \includegraphics[width=0.85\linewidth]{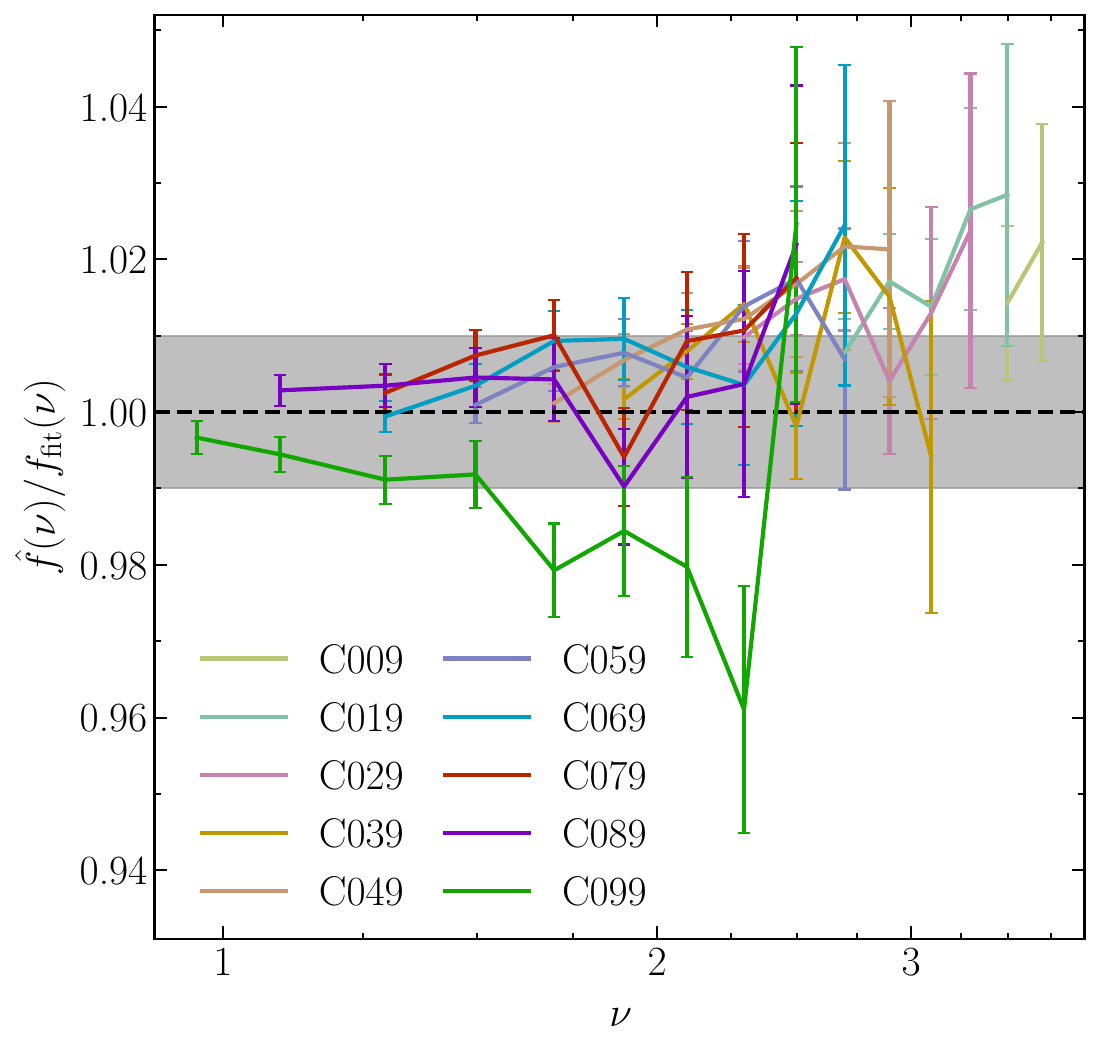}
    \caption{Ratios of $\hat{f}(\nu)$ measured in a subset of the AletheiaEmu simulations, with $\Delta = 200$, to the calibrated fitting function given in equation~(\ref{eq:fnu_ours}).}
    \label{fig:residuals_AletheiaEmu_m200b}
\end{figure}

Figure~\ref{fig:residuals_AletheiaEmu_m200b} shows the ratio of the measured $\hat{f}(\nu)$ from the AletheiaEmu simulations, with $\Delta = 200$, to our model $f_\mathrm{fit}(\nu)$. 
The prescription retains the accuracy shown for the Aletheia simulations (see Fig.~\ref{fig:residuals_aletheia_m200b}), with residuals nearly always within 3 per cent, across different shapes of the linear power spectrum and a wide range of clustering amplitudes, $\sigma_{12}$.

To provide an additional test of the reliability of our recipe, we compare its predictions to the HMF measured in the Uchuu simulation \citep{ishiyama_uchuu_dr1_2021}. 
Uchuu is a publicly available\footnote{\url{https://skiesanduniverses.iaa.es/Simulations/Uchuu/}} run produced with the \texttt{GreeM} code \citep{Ishiyama2009_GreeM1}. 
It features extremely high resolution ($12800^3$ particles in a $2\, h^{-1} \mathrm{Gpc}$ box) and assumes a $\Lambda\mathrm{CDM}$ cosmology based on \citet{planck_2015}, which is very similar to that of AletheiaMass and Aletheia Model 0. 
While its mass resolution is comparable to the $L_\mathrm{box} = 350 \, \mathrm{Mpc}$ AletheiaMass run, the much larger volume of probed by Uchuu allows for a significantly more precise measurement of the high-$\nu$ tail of the HMF compared to the simulations used for our calibration.

\begin{figure}
    \centering
    \includegraphics[width=0.95\linewidth]{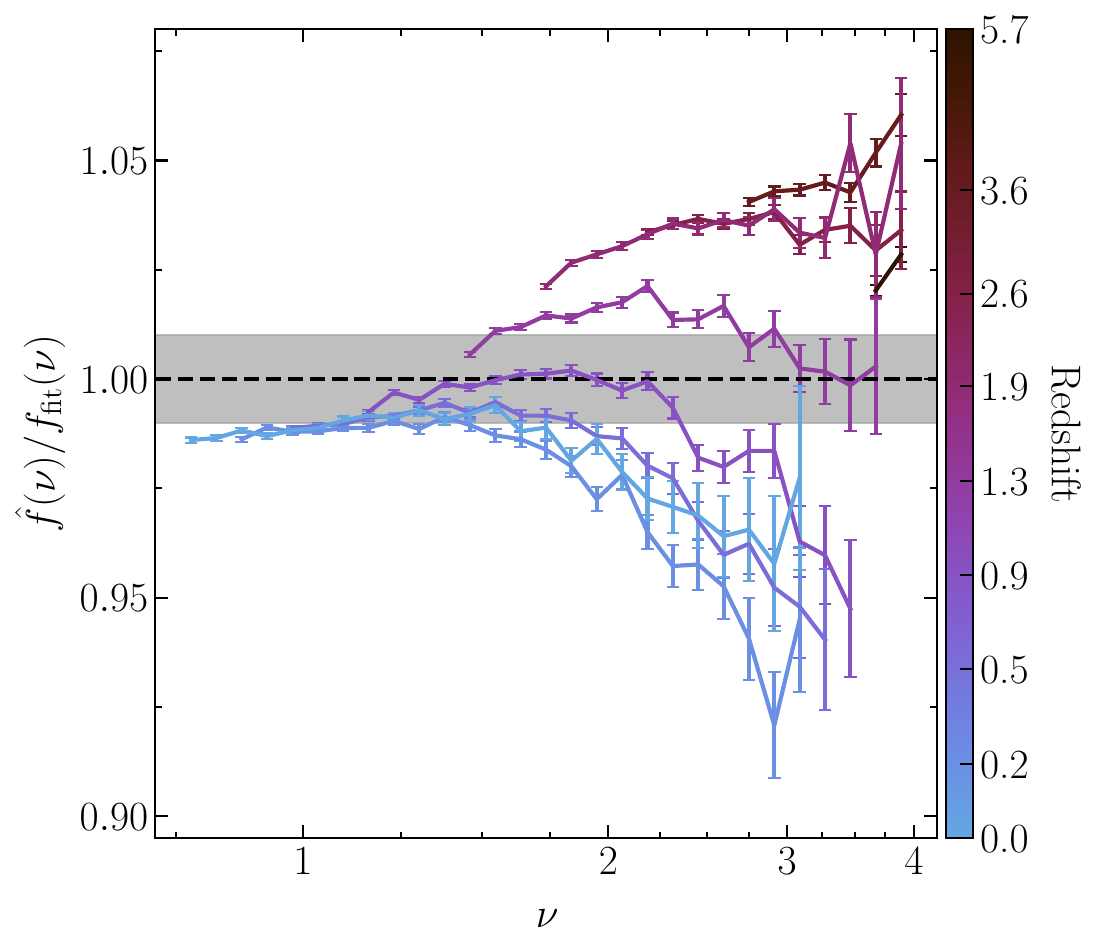}
    \caption{Ratios of $\hat{f}(\nu)$ measured in the Uchuu simulation to the calibrated fitting function given in equation~(\ref{eq:fnu_ours}), using parameter values interpolated to the corresponding $\Delta_\mathrm{vir}$ with equation~(\ref{eq:interpolation_fitting_parameters}). The grey band corresponds to a 1 per cent difference.}
    \label{fig:residuals_Uchuu_vir}
\end{figure}

The haloes of the Uchuu simulation were identified with \texttt{Rockstar} using the virial overdensity threshold, and the resulting catalogues are publicly available. For this analysis, we use a subset of 9 out of the 50 available snapshots, covering the redshift range $0 < z < 5.7$.
Figure~\ref{fig:residuals_Uchuu_vir} shows the ratio of our estimates of the halo multiplicity function derived from these samples and the predictions of our recipe, colour-coded by redshifts.

As shown in Fig.~\ref{fig:residuals_Uchuu_vir}, the accuracy of the model remains robust even on this independent dataset. 
The residuals are generally within 5 per cent, comparable to the precision obtained for the calibration suites, AletheiaMass (Fig.~\ref{fig:residuals_aletheiamass_vir}) and Aletheia (Fig.~\ref{fig:residuals_aletheia_vir}). 
A slight systematic trend in the ratios is visible, particularly for snapshots between $z = 1.90$ and $z = 0.19$. 
This behaviour is analogous to what we saw in the AletheiaMass virial HMF (see Fig.~\ref{fig:residuals_aletheiamass_vir}), but the much higher statistical precision of Uchuu makes this feature more evident. 
This confirms that the interpolation of parameters to the virial definition captures the dependence on the structure growth history slightly less precisely than a direct fit.
\begin{figure*}
    \includegraphics[width=0.9\linewidth]{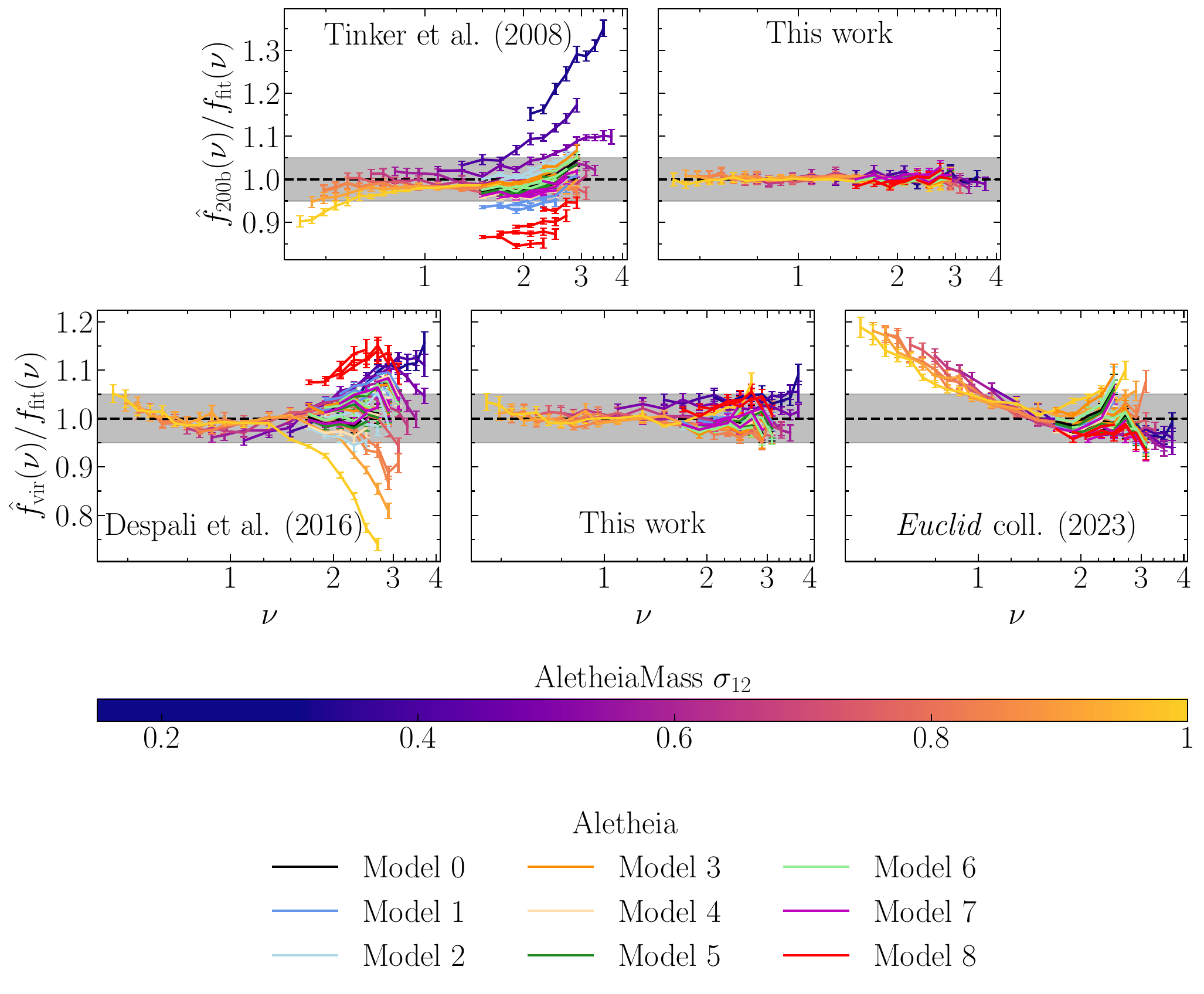}
    \caption{Ratios of $\hat{f}(\nu)$ from the Aletheia and AletheiaMass simulations, based on $\Delta = 200$ (upper panels) and $\Delta_\mathrm{vir}$ (lower panels), to other commonly used fitting functions \citep{tinker_hmf_2008, Despali_2016, castro_euclid_hmf_2023}, and the one introduced in this work. 
    The grey bands correspond to a 5 per cent difference.}
    \label{fig:comparison_other_recipes}
\end{figure*}

\subsection{Comparison with existing prescriptions}
\label{subsec:comparison}

After calibrating our model and demonstrating its accuracy, we compare its performance with some of the most commonly used HMF prescriptions.
We focus our comparison on measurements based on the two most commonly employed mass definitions, $M_{200\mathrm{b}}$ and $M_\mathrm{vir}$.
The results are shown in Fig.~\ref{fig:comparison_other_recipes}.

The upper panels, showing $\Delta = 200$ measurements, compare the accuracy of our model to that of the fitting function of \citet{tinker_hmf_2008}. 
While this prescription is remarkably accurate for a $\Lambda\mathrm{CDM}$ cosmology at $z \lesssim 1.5$, it loses accuracy at higher redshifts. 
Moreover, since it parametrises non-universality solely as a function of redshift, it does not account for differences arising from the specific structure formation history. 
Therefore, the ratios for the Aletheia simulations exhibit a spread comparable to that of the raw measurements shown in Fig.~\ref{fig:aletheia_ratio_tomodel0}, demonstrating that the model fails to capture this effect.

In the lower panels, we show the ratios of the virial $\hat{f}(\nu)$ to the recipes of \citet{Despali_2016} and \citet{castro_euclid_hmf_2023}.
For \citet{Despali_2016}, which assumes a universal multiplicity function, the residuals reach the 10 per cent level. 
While the dependency of the virial overdensity definition $\Delta_\mathrm{vir}$ on $\Omega_\mathrm{m}(z)$ (equation~\ref{eq:virial_delta}) partially absorbs the dependence on structure growth history, it is insufficient to capture the full time evolution of the HMF.

In contrast, the recipe of \citet{castro_euclid_hmf_2023} explicitly accounts for both structure formation history (via $\Omega_\mathrm{m}(z)$) and the power spectrum shape (via $\mathrm{d}\ln \sigma / \mathrm{d} \ln R$). 
This recipe achieves remarkable accuracy at $\nu \gtrsim 1.5$. 
However, a visible bias remains at lower $\nu$. 
This is likely because their model was calibrated specifically for the expected lower mass limit of the \textit{Euclid} galaxy cluster catalogue \citep[$M \gtrsim 10^{13}~h^{-1}~M_\odot$,][]{sartoris_euclid_clusters_2016}, leaving the low-mass end less constrained. 
Furthermore, the amplitude of the fitting function of \citet{castro_euclid_hmf_2023} was not fitted freely as the other parameters, but fixed to normalise the integral of $f_\mathrm{fit}(\nu)$.
Although this choice does not impact accuracy within their specific mass range of interest, it may contribute to degrade the performance of the fit at lower masses.

Finally, we emphasise that our model for the virial HMF is derived by interpolating parameters fitted to other mass definitions. 
In contrast, the functions of \citet{Despali_2016} and \citet{castro_euclid_hmf_2023} were directly calibrated on virial halo catalogues. 
Despite this disadvantage, our interpolated model offers a remarkably good performance.

\section{Conclusions}
\label{sec:conclusions}
The halo mass function constitutes a fundamental link between galaxy and cluster surveys and theoretical models of structure formation. In this paper, we calibrated a new semi-analytical prescription for the halo multiplicity function, $f(\nu)$, which lies at the core of any HMF model, owing to its near-universality once expressed as a function of the peak height $\nu$.

Our model is based on a physically motivated choice of parameters, as detailed in Section~\ref{sec:modelling}. 
In particular, we exploited the evolution mapping framework \citep{sanchez_evo_mapping_2022} to accurately account for the dependence of the HMF on the history of structure formation. 
We also took into account its dependence on the local shape of the linear matter power spectrum, following \citet{ondaromallea_hmf_2021} and \citet{castro_euclid_hmf_2023}.

We calibrated our model on measurements from two suites of simulations, Aletheia and AletheiaMass, the former representing cosmologies with different structure formation histories and the latter being a high-resolution $\Lambda\mathrm{CDM}$ suite. 
We extracted halo catalogues for ten different mass definitions, with overdensity thresholds ranging from 150 to 1600 times the background matter density setting the halo boundaries, refitting the parameters on each of them.

The calibrated fitting function for our reference mass definition, $\Delta = 200$, achieves per cent-level accuracy across the full range of masses, redshifts, and structure formation histories (see Figs.~\ref{fig:residuals_aletheiamass_m200b}~and~\ref{fig:residuals_aletheia_m200b}), and retains this performance when tested on cosmologies with different linear power spectrum shapes (see Fig.~\ref{fig:residuals_AletheiaEmu_m200b}).

Turning to other mass definitions, the accuracy remains robust, degrading only slightly to the 
5 per cent level at the highest values of $\Delta$. 
We note that at these high overdensities, the convergence of the simulation results is poorer, limiting the precision of the measurements available for calibration (see Fig.~\ref{fig:fit_allmasses}).

Furthermore, we provided fitting formulae to interpolate between the fitted parameters as a function of $\Delta$. 
We tested the resulting interpolated recipe on $\hat{f}(\nu)$ measured in virial halo catalogues and showed that its accuracy still lies well within 5 per cent.
We also compared, in Fig.~\ref{fig:comparison_other_recipes}, the accuracy of our model with that of other commonly used recipes \citep{tinker_hmf_2008, Despali_2016, castro_euclid_hmf_2023}. 
We found that our prescription yields competitive or superior accuracy across the full range of redshifts and cosmologies, successfully reproducing the detailed non-universal features of the HMF where other models show systematic deviations.

Future work building on the results of this paper includes building an emulator of the HMF informed by evolution mapping, exploiting the full set of the AletheiaEmu simulations \citep{sanchez_aletheia_2025}; incorporating the impact of massive neutrinos by leveraging the extended evolution mapping framework of \citet{pezzotta_evomapping_neutrinos_2025}; investigating other halo properties, such as bias, density profiles, and concentrations in light of evolution mapping; and modelling the abundances of spherical cosmic voids under the same theoretical lens.

In addition to providing a high-precision modelling tool for halo and galaxy clustering statistics, our results demonstrate the critical importance of the integrated growth of structure parameter, $\tilde{x}$, introduced by \citet{sanchez_aletheia_2025}, in characterizing deviations from universal evolution. 
By effectively capturing these non-universal features, our work highlights the potential and versatility of evolution mapping even in the highly non-linear regime. 
This framework simplifies the description of the matter density field, isolating the relevant physical quantities to offer a clearer intuition of structure formation.

\section*{Acknowledgements}

We would like to thank Andrea Pezzotta, Carlos Correa, Sofia Contarini, Agnė
Semėnaitė, Alejandro Pérez Fernández, Jiamin Hou, Dante Paz, Facundo Rodriguez,
Emiliano Sefusatti, Guido D'Amico, Sara Maleubre Molinero, Luca Fiorino,
Soumadeep Maiti and Fabian Balzer for their help and useful discussions. The
simulations used in this work, Aletheia, AletheiaMass and AletheiaEmu were
carried out and post-processed on the HPC systems Cobra, Raven and Viper of the
Max Planck Computing and Data Facility (MPCDF, \url{https://www.mpcdf.mpg.de})
in Garching, Germany. This project has received funding from the European
Union’s HORIZON-MSCA-2021-SE-01 Research and Innovation programme under the
Marie Skłodowska-Curie grant agreement number 101086388 - Project acronym:
LACEGAL.

\section*{Data Availability}

The simulation data underlying this work will be shared upon reasonable request to the corresponding authors.


\bibliographystyle{mnras}
\bibliography{bibliography}



\appendix

\section{Calibration for spherical overdensity haloes}
\label{appendix_strict_SO}

\begin{figure*}
    \centering
    \includegraphics[width=0.9\textwidth]{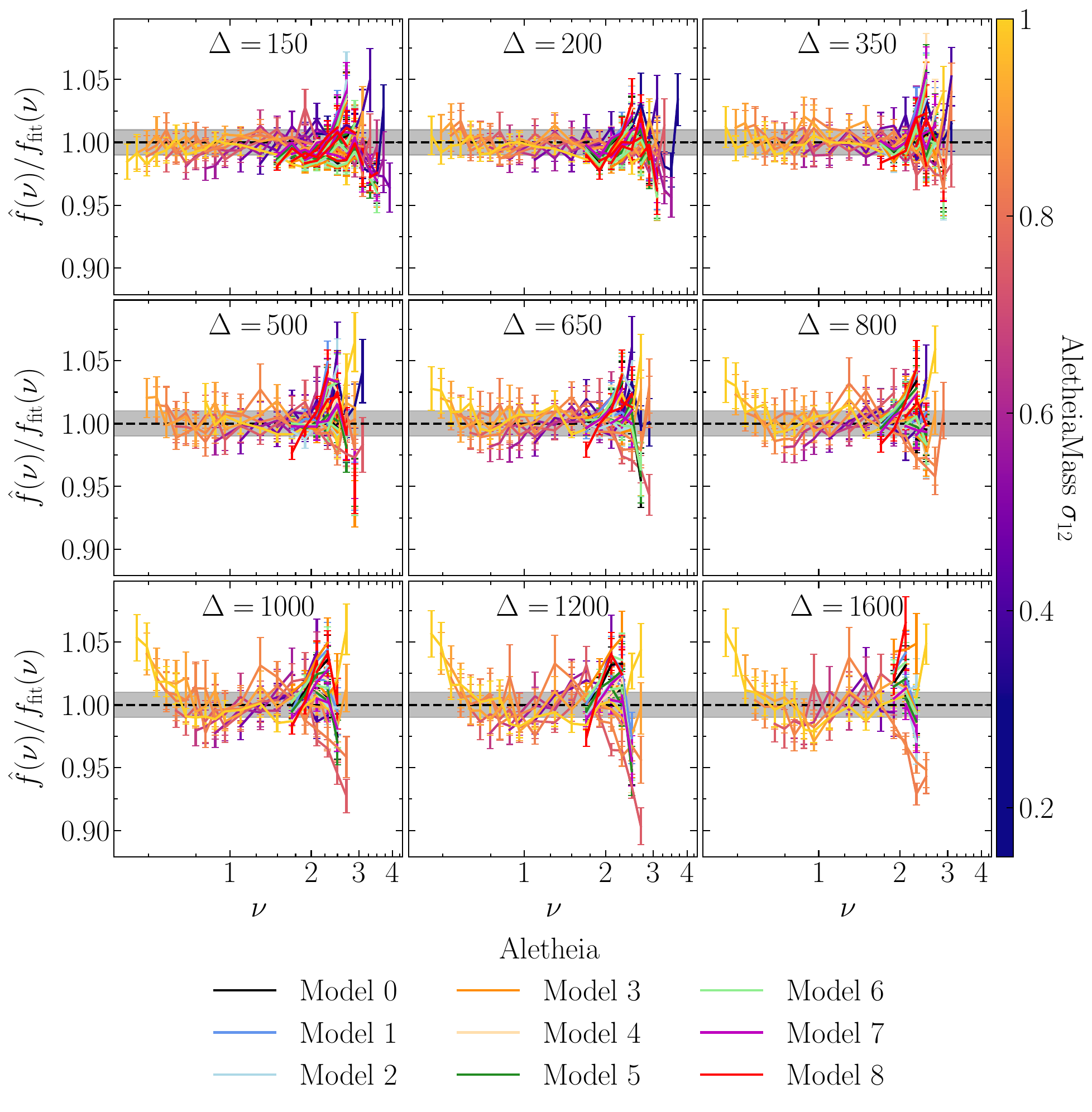}
    \caption{Ratios of $\hat{f}(\nu)$ measured in the Aletheia and AletheiaMass simulations to the fitting function given in equation~(\ref{eq:fnu_ours}), for  different overdensity thresholds $\Delta$, including unbound particles. 
    The grey bands correspond to a 1 per cent difference.}
    \label{fig:fit_allmasses_SO}
\end{figure*}

Here, we provide a second set of fitted parameters, most suited to model the HMF derived form spherical overdensity halo finders. 
Specifically, we use the same simulations and the same measurement pipeline as described in Section~\ref{sec:method}, but with the flag \texttt{STRICT\_SO\_MASSES} set to 1 in the \texttt{Rockstar} configuration files. This prevents the code from excluding particles that are not gravitationally bound to the haloes.
We report the fitted parameters for nine different mass definitions in Table~\ref{tab:fitting_parameters_SO}.
\begin{table*}
    \centering
    \caption{Best-fitting parameters for our model multiplicity function, equation~ (\ref{eq:fnu_ours}), for different overdensity thresholds, for halo masses including the contribution of unbound particles.}
    \begin{tabularx}{\textwidth}{Y|YYYYYYYYY}
        \hline
         $\Delta$ & $A_0$ & $A_n$ & $a_n$ & $B_0$ & $B_x$ & $B_n$ & $b_n$ & $C$ & $\log_{10} \eta$ \\
         \hline
        150 & 0.33 & -0.11 & 0.23 & -0.55 & 1.35 & 0.22 & -0.55 & 0.418 & -0.50  \\
        200 & 0.46 & 0.04 & 0.210 & -0.55 & 0.94 & 0.10 & -0.66 & 0.438 & -0.62  \\
        350 & 0.62 & 0.15 & 0.208 & -0.55 & 0.69 & 0.029 & -0.79 & 0.481 & -0.77  \\
        500 & 0.67 & 0.18 & 0.203 & -0.55 & 0.62 & 0.007 & -0.87 & 0.52 & -0.95  \\
        650 & 0.70 & 0.194 & 0.211 & -0.55 & 0.58 & -0.004 & -0.9 & 0.547 & -1.09  \\
        800 & 0.76 & 0.22 & 0.209 & -0.55 & 0.52 & -0.025 & -0.91 & 0.56 & -1.22  \\
        1000 & 0.82 & 0.246 & 0.190 & -0.55 & 0.46 & -0.047 & -0.89 & 0.58 & -1.58  \\
        1200 & 0.88 & 0.267 & 0.188 & -0.55 & 0.43 & -0.059 & -0.88 & 0.59 & -1.90  \\
        1600 & 1.00 & 0.30 & 0.190 & -0.55 & 0.37 & -0.08 & -0.93 & 0.63 & $< -2.5$  \\
         \hline
    \end{tabularx}
    \label{tab:fitting_parameters_SO}
\end{table*}
In Fig.~\ref{fig:fit_allmasses_SO}, we show the ratios of $\hat{f}(\nu)$ from the Aletheia and AletheiaMass simulations. The model achieves comparable accuracy to the recipe given in the main text, slowly worsening as $\Delta$ increases but always within 5 per cent, even at the highest overdensity thresholds.

Moreover, we provide a set of empirical interpolating curves to allow the computation of the model on any $\Delta \in [150,~1600]$:
\begin{equation}
    \begin{split}
        &A_0 = 0.45\, B_x^{-0.81} \, , \\
        &A_n = -0.44 \, B_x + 0.453 \, , \\
        &a_n = -0.038 y + 0.31 \, , \\
        &B_x = -0.7 y^2 + 3.8 y -4.9 + \frac{0.25}{y}  \, , \\
        &B_n =  0.31 \, B_x - 0.189 \, , \\
        &b_n = -0.93 \, {(y - 2.13)}^{0.18} \, , \\
        &C = 0.0382 y^2 + 0.238 \, , \\
        &\log_{10} \eta = -1.4 \, {\Big(\frac{\Delta}{1000}\Big)}^{0.54} \, . 
    \label{eq:interpolation_fitting_parameters_SO}
    \end{split}
\end{equation}

\begin{figure*}
    \centering
    \includegraphics[width=0.9\textwidth]{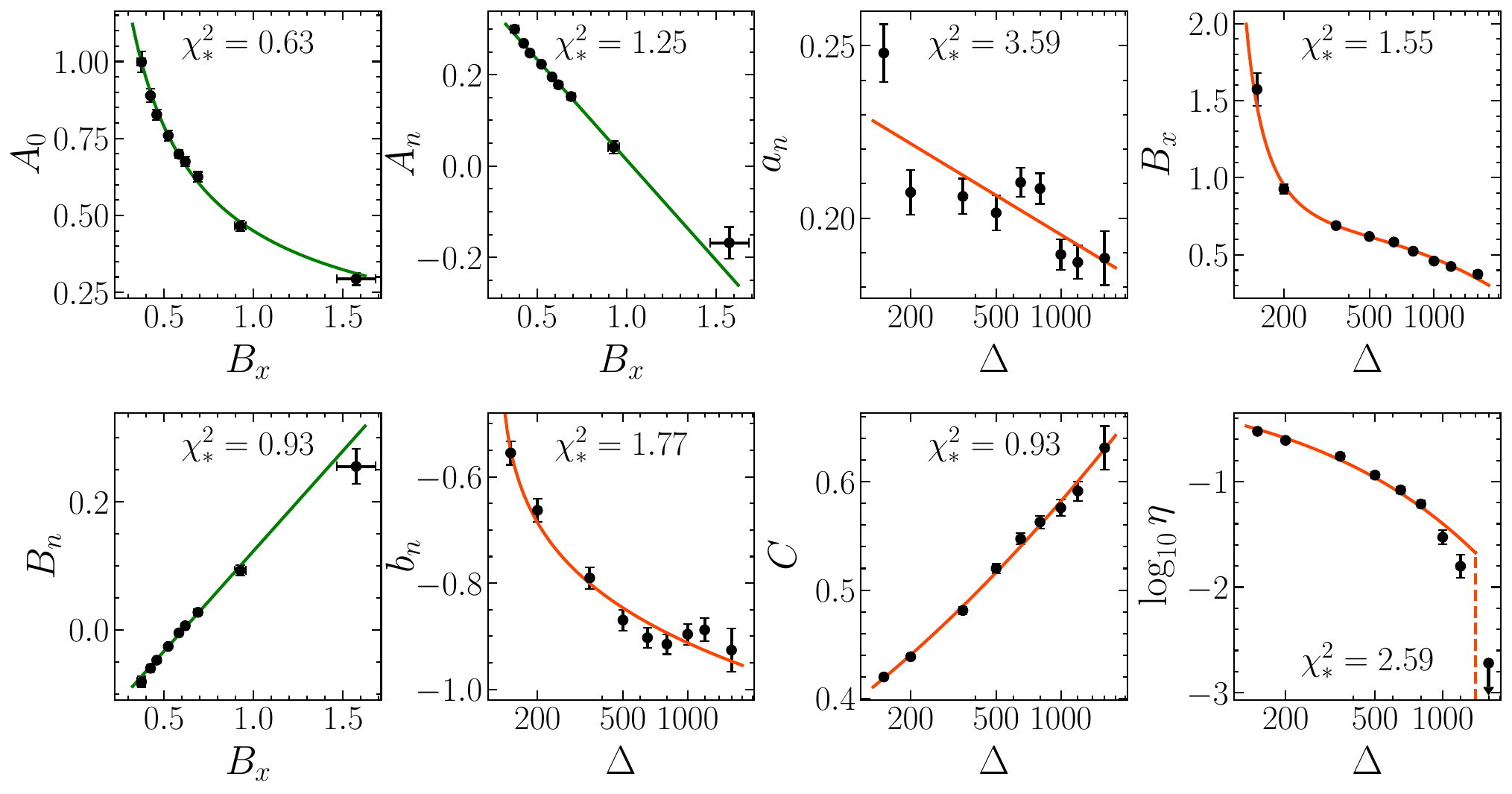}
    
    \caption{Interpolation of the best-fitting parameters for halo masses including the contribution of unbound particles. The black points represent the values listed in Table~\ref{tab:fitting_parameters_SO}, while lines represent the empirical interpolating functions given in equation~(\ref{eq:interpolation_fitting_parameters_SO}). The fitting function for the memory parameter $\eta$ is truncated at $\Delta = 1400$, above which we set $\eta = 0$ (see Section~\ref{subsec:fit_all}).
    Interpolations as a function of $y=\log_{10}\Delta$ are shown as red lines, while the ones given as a function of $B_x$ are shown in green. In each panel, we report the reduced chi-square of the respective fit, $\chi^2_* = \chi^2 / \mathrm{d.o.f.}$}
    \label{fig:interpolation_fitting_parameters_SO}
\end{figure*}

We show the interpolation of the fitted parameters in Fig.~\ref{fig:interpolation_fitting_parameters_SO}.


\bsp	
\label{lastpage}
\end{document}